\documentclass{elsart}
\usepackage{graphicx}
\usepackage{epsfig}

\newcommand{\D}{\displaystyle}

\def\bs{\bigskip}

\def\ea{et al.\,}
\def\eg{{\it e.g.\,}}
\def\be{\begin{equation}}
\def\ee{\end{equation}}
\def\apj{ApJ}

\def\mn{MNRAS}

\def\nrel{nonrelativistic\,\,}

\begin{document}
\begin{frontmatter} 

\title{Modeling Integrated Properties and the Polarization of the 
Sunyaev-Zeldovich Effect}

\author[tau,ucsd]{Y.~Rephaeli}, \author[tau]{S.~Sadeh} \& 
\author[ucsd]{M.~Shimon} 
\address[tau]{School of Physics \& Astronomy, Tel Aviv University, 
Tel Aviv, Israel}
\address[ucsd]{Center for Astrophysics and Space Sciences, UCSD, 
La Jolla, CA, USA}
\maketitle

\begin{abstract}

Two little explored aspects of Compton scattering of the CMB in 
clusters are discussed: The statistical properties of the 
Sunyaev-Zeldovich (S-Z) effect in the context of a non-Gaussian 
density fluctuation field, and the polarization patterns in a 
hydrodynamcially-simulated cluster. We have calculated and compared 
the power spectrum and cluster number counts predicted within the 
framework of two density fields that yield different cluster mass 
functions at high redshifts. This is done for the usual Press \& 
Schechter mass function, which is based on a Gaussian density 
fluctuation field, and for a mass function based on a 
$\chi^2$-distributed density field. We quantify the significant 
differences in the respective integrated S-Z observables in these 
two models. 

S-Z polarization levels and patterns strongly depend on 
the non-uniform distributions of intracluster gas and on peculiar 
and internal velocities. We have therefore calculated the patterns 
of two polarization components that are produced when the CMB is 
doubly scattered in a simulated cluster. These are found to be 
very different than the patterns calculated based on spherical 
clusters with uniform structure and simplified gas distribution.

\end{abstract}


\end{frontmatter}

\section{Introduction}

The pioneering experimental and observational CMB work of our esteemed 
colleague Francesco Melchiorri has paved the way for generations of 
students and colleagues in Italy and elsewhere, who are continuing the 
mission he helped define and lead for more than 30 years. Francesco 
worked hard to advance experimental and observational capabilities to 
measure the Sunyaev-Zeldovich (S-Z) effect in nearby clusters. His 
expertise and leadership resulted in the successful development of 
MITO, the measurement of the effect in the Coma cluster (De Petris 
\ea 2002), and the upgrading of MITO to a bolometer array (Lamagna 
\ea 2005). As noted by Rephaeli (2005) in Francesco's obituary, his 
passing marks the end of the founder's era in the growth of the 
thriving Italian CMB community.

In light of current and future capabilities of the many upcoming 
S-Z experiments, we have calculated the power spectrum and cluster 
number counts in the context of a specific non-Gaussian density 
fluctuation field. As we discuss in the next section, certain 
observational results point to the possible need for excess power 
on cluster scales. The enhanced interest in non-Gaussian models 
motivates contrasting the integrated S-Z observables of one such 
viable model with those of the standard Gaussian model whose parameters 
are well determined.

Future sensitive mapping of the S-Z effect in individual clusters 
will allow measuring its polarization level. A basic description 
of some of the S-Z polarization patterns in clusters has already 
been given by Sazonov \& Sunyaev (1999). In Section 3 we briefly 
summarize some of the features of two of the polarization components 
that are induced when the CMB is doubly scattered in a (non-idealized) 
cluster whose dynamical state and gas properties are deduced directly 
from hydrodynamical simulations.

\section{Integrated S-Z Observables in a Non-Gaussian Model}

It has been suggested (Mathis, Diego, \& Silk 2004) that some 
observational results possibly point to an appreciable deviation from a 
Gaussian probability distribution function (PDF) of the primordial 
density fluctuation field. These are the relatively high CMB power at 
high multipoles measured by the \textit{CBI} (Mason et al. 2003) and 
\textit{ACBAR} (Kuo et al. 2004) experiments, the detection of 
structures with high velocity dispersions at redshifts $z=4.1$ (Miley 
et al. 2004) and $z=2.1$ (Kurk et al. 2004), and the apparent slow 
evolution of the X-ray luminosity function (as deduced from catalogs 
compiled by Vikhlinin et al. 1998 and Mullis et al. 2003).

Analyses of \textit{WMAP} all-sky maps (Komatsu et al. 2003, McEwen 
et al. 2006) do not rule out an appreciable non-Gaussian PDF tail on 
clusters scales. Enhanced power on the scales of clusters may lead to 
detectable differences in S-Z power levels and cluster number counts 
as compared with those predicted from the standard Gaussian PDF. It is 
of interest, therefore, to explore the consequences of a non-Gaussian 
PDF, especially in light of the large S-Z surveys that will be conducted 
by ground-based telescopes and the Planck satellite. These will measure 
the effect in thousands of clusters and map the anisotropy it induces in 
the CMB spatial structure. 

The non-Gaussian model we consider here has its origin in isocurvature 
fluctuations produced by massive scalar fields that are thought to have 
been present during inflation. The model is characterized by a PDF that 
has a $\chi^2_m$ distribution of $m$ CDM fields which are added in 
quadratures (Peebles 1997, 1999, Koyama, Soda \& Taruya 1999).

We have carried out detailed calculations of the S-Z power spectrum and 
cluster number counts predicted in the standard Gaussian and a 
non-Gaussian model with $m=1, 2$. Below we briefly describe the 
principal aspects of our work and its main results; a more comprehensive 
description can be found in Sadeh, Rephaeli \& Silk (2006; hereafter, 
SRS).

\subsection{Calculations}

The integrated statistical properties of the population of clusters 
involve the basic cosmological and large-scale quantities, and essential 
properties of intracluster (IC) gas. For Gaussian models these properties 
have been quantitatively determined by semi-analytic calculations (e.g., 
Kaiser 1981, Colafrancesco \ea 1994, 1997, Molnar \& Birkinshaw 2000, 
Sadeh \& Rephaeli 2004) and from hydrodynamical simulations (e.g., 
Springel \ea 2001). 

The Press \& Schechter (1974) mass function is 
\begin{equation}
n(M,z)=-F(\mu)\frac{\rho_b}{M\sigma}\frac{d\sigma}{dM}dM,
\label{eq:nmz}
\end{equation}
where $\mu\equiv\delta_c/\sigma$ and $\rho_b$ is the background 
density. The critical overdensity for spherical collapse (which is only 
weakly dependent on redshift) is assumed to be constant, $\delta_c=1.69$, 
and the density field variance, smoothed over a top-hat window function of 
radius $R$, is 
\begin{equation}
\sigma^{2}(R)=\int_0^{\infty}P(k)\widetilde{W}^2(kR)k^2dk,
\end{equation}
where $P(k)\equiv Ak^{n}T^{2}(k)$. Evolution of the mass variance 
is given in
\begin{equation}
\sigma(M,z)=\frac{g[\Omega_m(z)]}{g[\Omega_m(0)]}
\frac{\sigma(M,0)}{1+z},
\end{equation}
where
\begin{equation}
\Omega_m(z)=\frac{\Omega_m(0)(1+z)^3}{(1+z)^2[1+\Omega_m(0)z-
\Omega_{\Lambda}(0)]}; \,\,\,\Omega_{\Lambda}(z)=1-\Omega_m(z),
\end{equation}
and (Carroll, Press \& Turner 1992)
\begin{equation}
g[\Omega_m(z)]=\frac{2.5\Omega_m(z)}{[\Omega_m(z)^{4/7}-\Omega_{\Lambda}+
(1+\Omega_m(z)/2)(1+\Omega_{\Lambda}/70)]}. 
\end{equation}
The function $F(\mu)$ in equation~(\ref{eq:nmz}) assumes the form 
\begin{equation}
F(\mu)=\sqrt{\frac{2}{\pi}}\D{e^{-\frac{\mu^2}{2}}}\frac{\mu}{\sigma_R},
\label{eq:f_g}
\end{equation}
for a Gaussian density field, and the form 
\begin{equation}
F(\mu)=\frac{\left(1+\sqrt{\frac{2}{m}}\mu\right)^{m/2-1}}{(\frac{2}{m})
^{(m-1)/2}\Gamma{(\frac{m}{2})}}e^{\left[-\frac{m}{2}\left(1+\sqrt
{\frac{m}{2}}\mu\right)\right]}\frac{\mu}{\sigma_R},
\label{eq:f_ch}
\end{equation}
for a density fluctuation field which is distributed according to the
$\chi^2_m$ model mentioned above.

The highest contrast with the standard model is obtained for $m=1$, 
the case we have mostly explored and discuss here, but in order to 
demonstrate the impact of this parameter on PDF tail, we have also 
considered the case $m=2$ (for details, see SRS). Note that the 
higher the value of $m$, the closer is the PDF to a Gaussian, as 
would be expected from the central limit theorem. 

The CDM transfer function for the Gaussian model is 
\begin{equation}
T(k)=\frac{\ln{(1+2.34q)}}{2.34q}[1+3.89q+(16.1q)^2+(5.46q)^3+
(6.71q)^4]^{-1/4}.
\end{equation}
For the the $\chi^{2}_m$ model, we adopt the isocurvature CDM transfer 
function 
\begin{equation}
T(k)_{iso}=(5.6q)^{2}\left[1+\frac{(40q)^2}{1+
215q+(16q)^2/(1+0.5q)} + (5.6q)^{8/5}\right]^{-5/4},
\end{equation}
with $q\equiv k/(\Omega_m h^2) \, Mpc^{-1}$ (Bardeen et al. 1986). 
We adopt the (now) standard $\Lambda$CDM flat cosmological model, with 
$\Omega_{\Lambda}=0.7$, $\Omega_m=0.3$, $h=0.7$. In the Gaussian 
case the spectral index was taken to be $n=1$ with mass variance 
normalization $\sigma_8=0.9$. In the non-Gaussian model $n=-1.8$, and 
by requiring that the present cluster abundance is the same as calculated 
in the Gaussian model, the value $\sigma_8=0.73$ was deduced the 
$\chi^{2}_1$ model.

The basic expression in the calculation of the S-Z angular power 
spectrum is 
\begin{equation}
C_{\ell}=\int_z\,r^2\frac{dr}{dz}\int_M n(M,z)\,G_{\ell}(M,z)\,dM\,dz,
\label{eq:clmb}
\end{equation}
where $r$ is the co-moving radial distance, and $G_{\ell}$ is obtained 
from the angular Fourier transform of the profile of the thermal S-Z 
temperature change at an angular distance $\theta$ from the center of 
a cluster, , $\Delta T(\theta)$. The function $G_{\ell}$, which is 
proportional to $(\Delta T)^2$, is fully specified in, e.g., Molnar 
\& Birkinshaw (2000). In the limit of non-relativistic electron 
velocities the relative temperature change assumes the simple form 
\begin{equation}
\frac{\Delta T}{T}(\theta)=\left[x\coth{\left(\frac{x}{2}\right)}-4\right]
y(\theta).
\end{equation}  
For an isothermal cluster with a $\beta$ density profile, the Comptonization 
parameter is 
\begin{eqnarray}
y(\theta)=\frac{2k_B\sigma_T}{m_e c^2} \frac{n_0(M,z)T_0(M,z)r_c(M,z)} 
{\sqrt{1+(\theta/\theta_c)^2}}\tan^{-1}\left[p\sqrt{\frac{1-
(\theta/p\theta_c)^2}{1+(\theta/\theta_c)^2}}\right] ,
\label{eq:ythet}
\end{eqnarray}
where $n_0$, $T_0$, are the central electron density, temperature, and 
$r_c$, and $\theta_c$ are the core radius and the angle subtended by 
the core, respectively. The gas profile is truncated at an outer radius 
$R=10r_c$.

The mass-temperature scaling relation is of basic importance for relating 
the magnitude of the S-Z effect to the cluster mass, and thereby to the 
mass function. We have used the following form for this relation 
\begin{equation}
T(M,z)=T_{15}\left(\frac{M}{10^{15}h^{-1}M_{\odot}}\right)^{\alpha}
(1+z)^{\psi},
\label{eq:tempsc}
\end{equation}
normalized such that the gas temperature of a $10^{15} h^{-1}M_{\odot}$ 
cluster is $8.5\,keV$ (at $z=0$). The parameter $\alpha$ is usually 
taken to be $2/3$, in accordance with theoretical predictions based on 
hydrostatic equilibrium; we use a range of values that $\pm 10\%$ around 
$2/3$. The uncertainty in the evolution of the gas temperature is 
accounted for by taking the two extreme values of $0$ and $1$ for the 
parameter $\psi$, signifying either no evolution, or strong 
evolution, respectively. 

The electron density is 
\begin{equation}
n_e(M,z)\simeq n_0\frac{f}{0.1}(1+z)^{3}, 
\end{equation}
with the scaling of the gas mass fraction to $\sim 10\%$ (\eg, Carlstrom, 
Holder \& Reese 2002). Since we do not yet know the redshift dependence of 
$f$, we assume $f=0.1$ to be roughly valid throughout the redshift interval 
($0-6$) considered here. Finally, the core radius is calibrated according 
to the simple relation
\begin{equation}
r_c=0.15h^{-1}\,\left(\frac{M}{10^{15}h^{-1}M_{\odot}}\right)^{1/3}\frac{1}{1+z} \,Mpc.
\end{equation}
Observations indicate a variance of $\sim 20\%$ in the 
value of the core radius, which is assumed here, with a corresponding 
range in the value of the central density.

The number of clusters with S-Z flux (change) above 
$\Delta \overline{F}_{\nu}$ is (e.g. Colafrancesco et al. 1997)
\begin{equation}
N(>\Delta \overline{F}_{\nu})=\int r^{2}\frac{dr}{dz}dz 
\int_{\Delta \overline{F}_{\nu}}n(M,z)\,dM.
\label{eq:nc}
\end{equation}
For a cluster with mass M at redshift z 
\begin{equation}\Delta F_{\nu}=\D\frac{2(k_{B}T)^{3}}{(hc)^{2}}g(x)y_{0}
\int R_{s}(|\hat{\gamma}-\hat{\gamma_{\ell}}|,\sigma_{B})\cdot
y(|\hat{\gamma_{\ell}}|,M,z)\,d\Omega,
\end{equation}
where $\hat{\gamma_{\ell}}$ and $\hat{\gamma}$ denote line of sight 
(los) directions through the cluster center, and relative to this central los, 
respectively, 
\begin{equation}
y_0=2\frac{k_B\sigma_T}{m_e c^2}n_0(M,z)T_0(M,z)r_c(M,z),
\end{equation}
and the spectral dependence (of the thermal component) of the effect is 
given (in the \nrel limit; more on this later in this section) by 
\begin{equation}
g(x)\equiv\frac{x^4e^{x}}{(e^{x}-1)^2}\left[x\coth{(x/2)-4}\right] , 
\end{equation}
where $x=h\nu/kT$.
The profile of the effect is given in 
\begin{equation}
y(|\hat{\gamma_{\ell}}|,M,z)\equiv\D\frac{1}{\sqrt{1+(\theta/\theta_{c})^{2}}}
\cdot\tan^{-1}
{\left[p\sqrt{\frac{1-(\theta/p\theta_{c})^{2}}{1+(\theta/\theta_{c})^{2}}}\right]},
\end{equation}
which is the los integral along a direction that forms an angle $\theta$ 
with the cluster center. $R_s(|\hat{\gamma}-\hat{\gamma_{\ell}}|,
\sigma_{B})$ denotes the angular response of a detector whose beam size 
is given in terms of $\sigma_B$. Finally, 
\begin{equation}
\Delta\overline{F}_{\nu}=\D\frac{\int \Delta F_{\nu}E(\nu)d\nu}{\int E(\nu)d\nu},
\end{equation}
is the flux weighted over the detector spectral response $E(\nu)$. 

For determining S-Z cluster number counts we adopt the \textit{PLANCK/HFI} 
detection limit of $30\,mJy$. This flux limit translates to a lower limit 
of the integral over the mass function, equation~(\ref{eq:nc}). The 
results for the S-Z power spectrum and number counts described below refer 
to these two models: (a) IC gas temperature evolves with time according to 
$\psi=1$, and (b) no temperature evolution, $\psi=0$. Power spectra 
were calculated at $\nu =353$ GHz, with a $7.1'$ beam size. Number counts 
were calculated also at $\nu=143$ GHz and $\nu=545$ GHz.

\subsection{Results}

The augmented tail of the non-Gaussian distribution function (compared 
with that of a Gaussian) leads to earlier cluster formation (by virtue 
of the higher probability that an overdense region attains the critical 
density for collapse) and enhanced high-mass cluster abundance. Thus, in 
the non-Gaussian case the S-Z power spectrum reaches higher levels, and 
its peak shifts to higher multipoles, reflecting the higher density of 
distant clusters. This effect is greatly reduced in a non-evolving 
temperature model (in which the $1+z$ scaling is factored out). 

\begin{figure*}[t]
\centering
\vspace{9pt}
\epsfxsize=2.5in
\includegraphics[width=6.5cm,height=6cm]{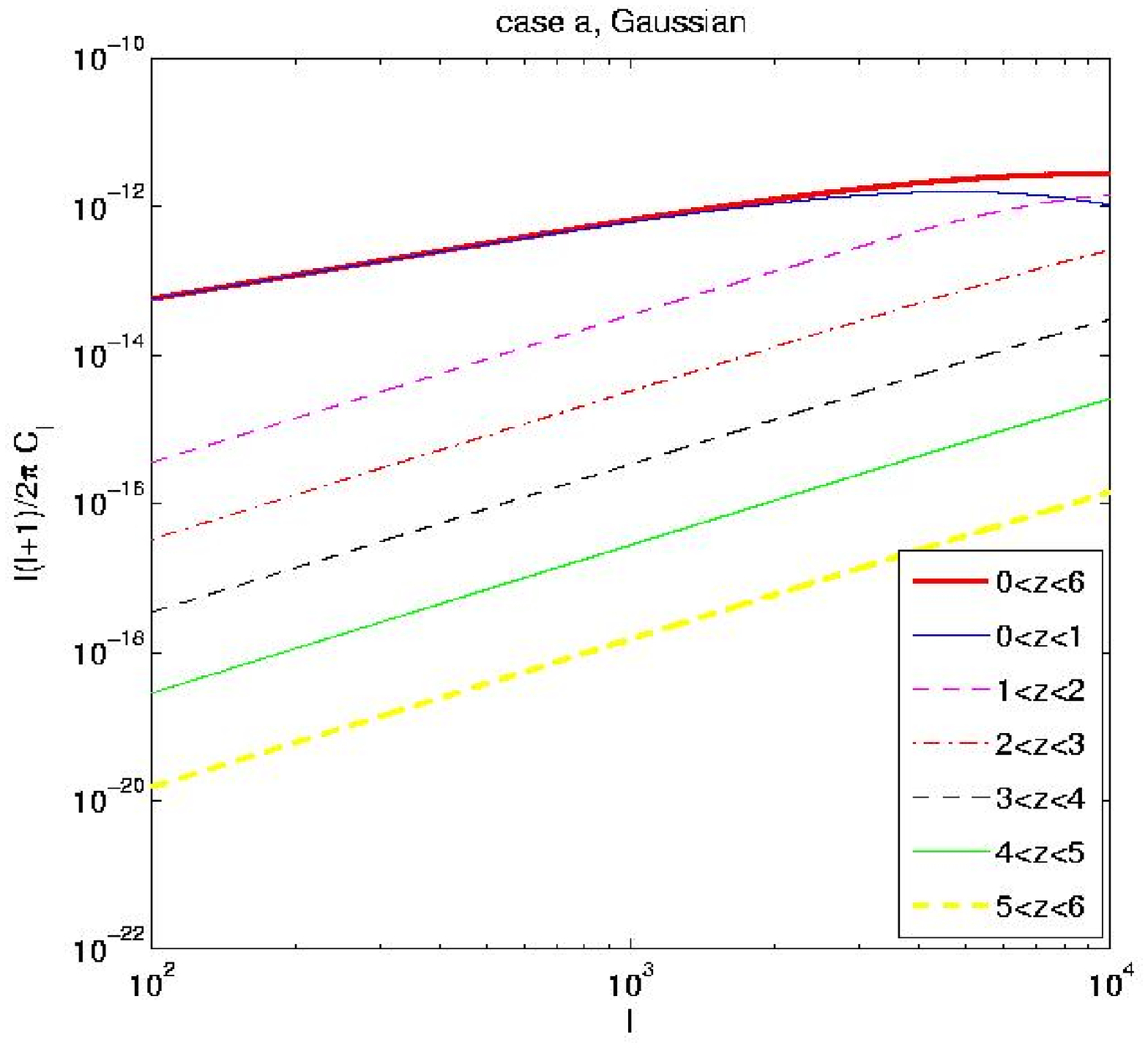}
\hspace{0.1in}
\includegraphics[width=6.5cm,height=6cm]{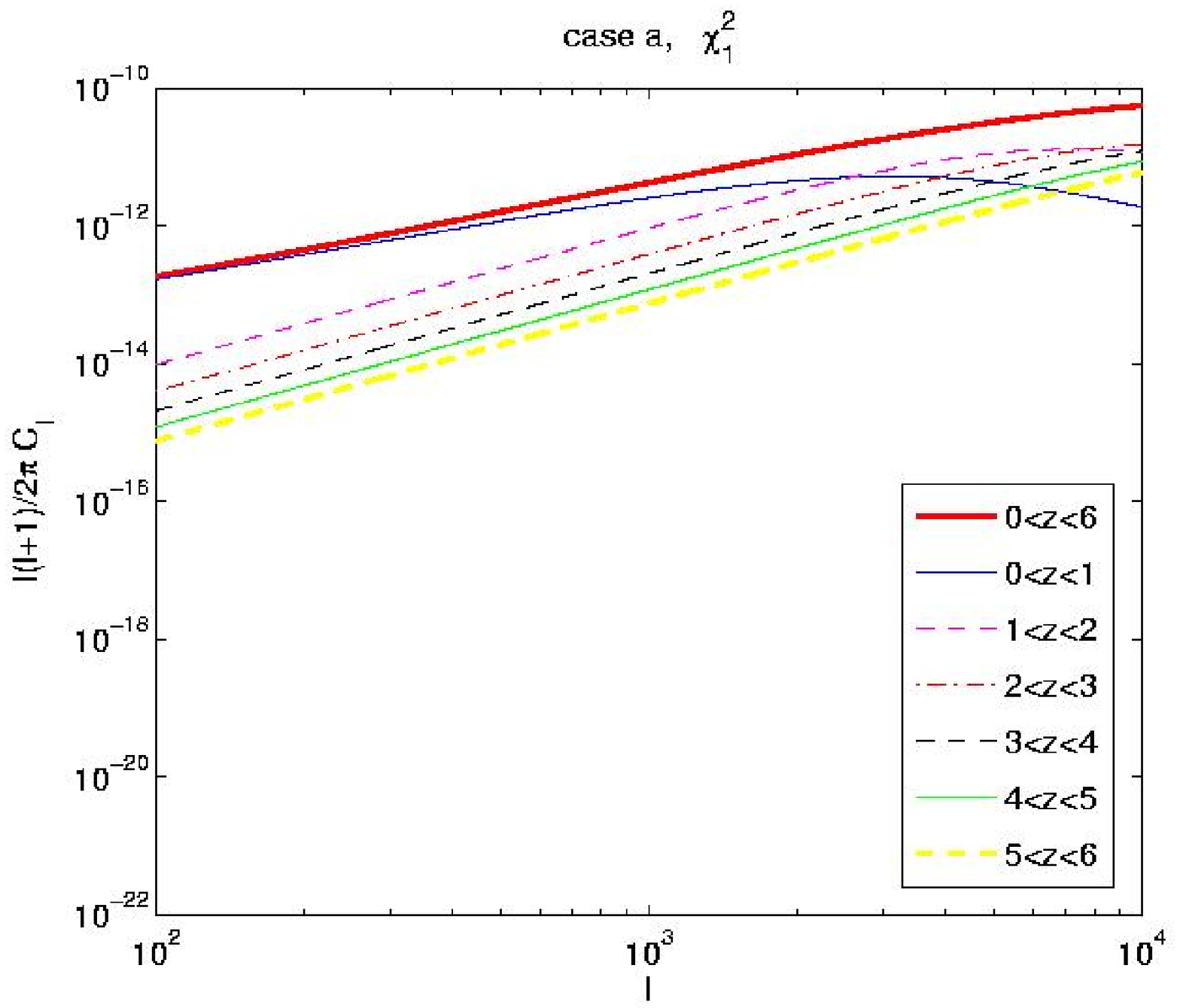}
\vspace{9pt}
\epsfxsize=2.5in
\includegraphics[width=6.5cm,height=6cm]{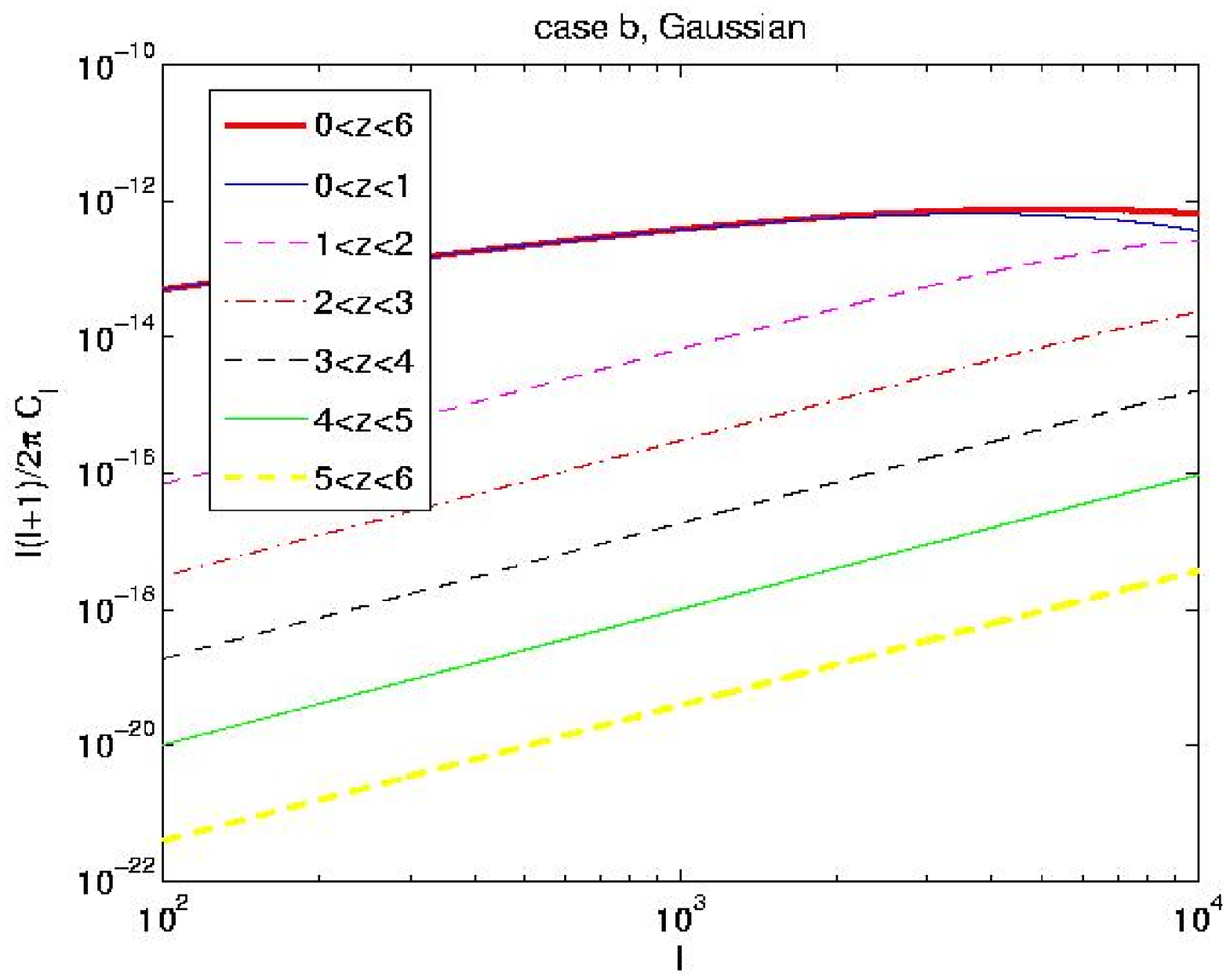}
\hspace{0.1in}
\includegraphics[width=6.5cm,height=6cm]{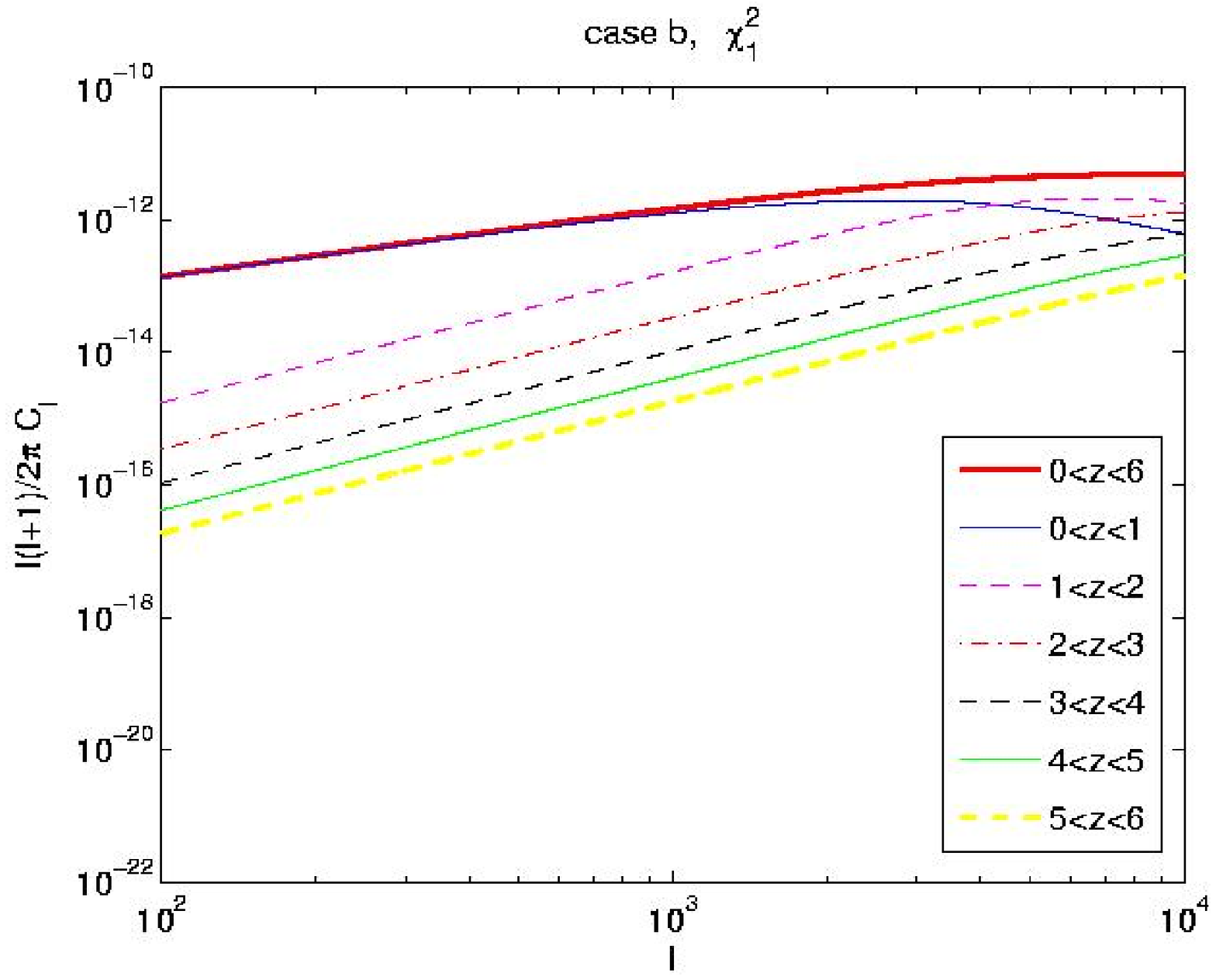}
\caption{Power spectrum of the S-Z effect. Shown are levels of power 
contributed by clusters in the redshift range of $0<z<6$, and at 6 redshift 
intervals. Left and right-hand panels correspond to the Gaussian and 
$\chi^{2}_1$ models, respectively. Upper and lower panels relate to cases 
(a) and (b), respectively}.
\label{fig:clz}
\end{figure*}

\begin{figure*}
\centering
\vspace{9pt}
\epsfxsize=2.5in
\includegraphics[width=6.5cm,height=6cm]{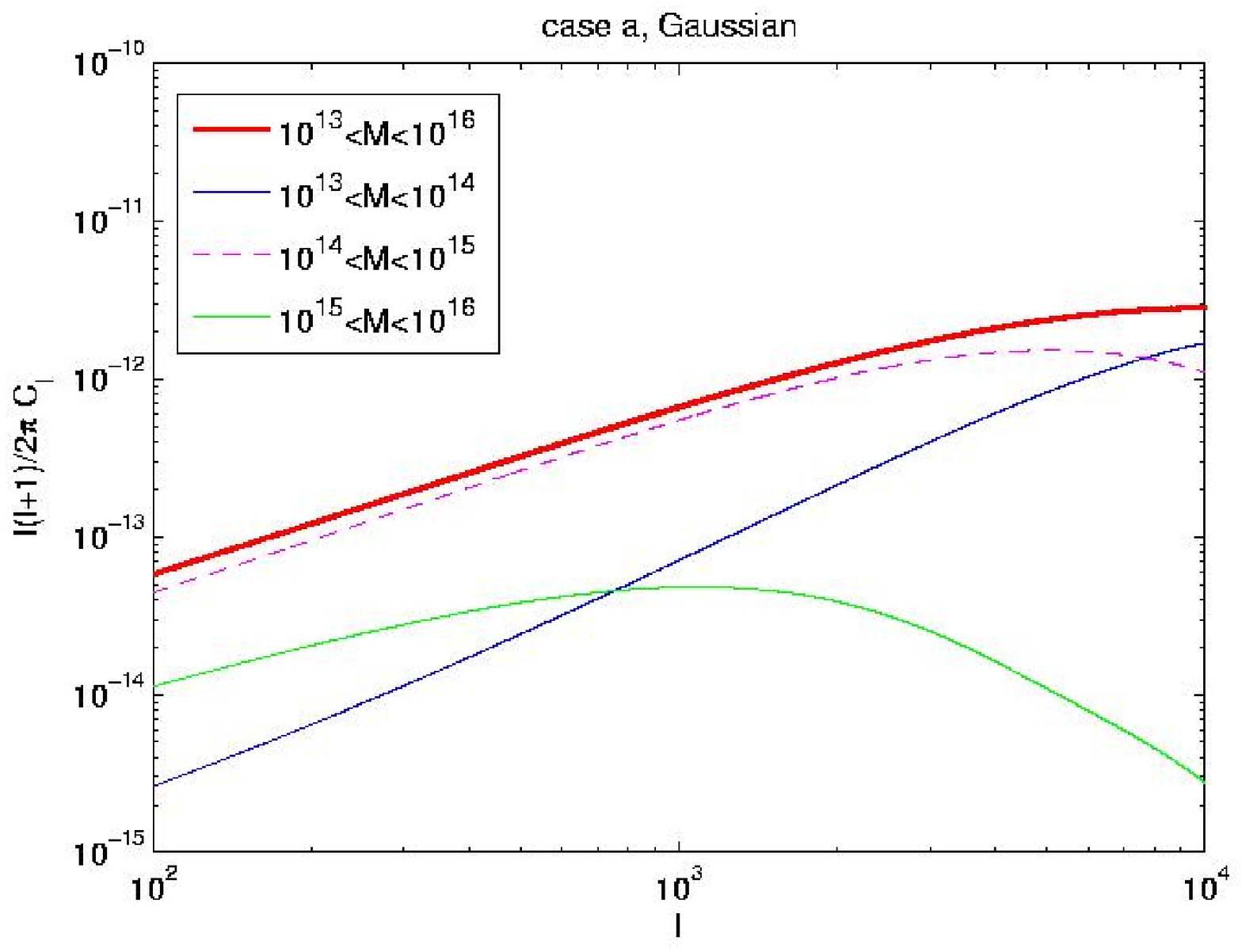}
\hspace{0.1in}
\includegraphics[width=6.5cm,height=6cm]{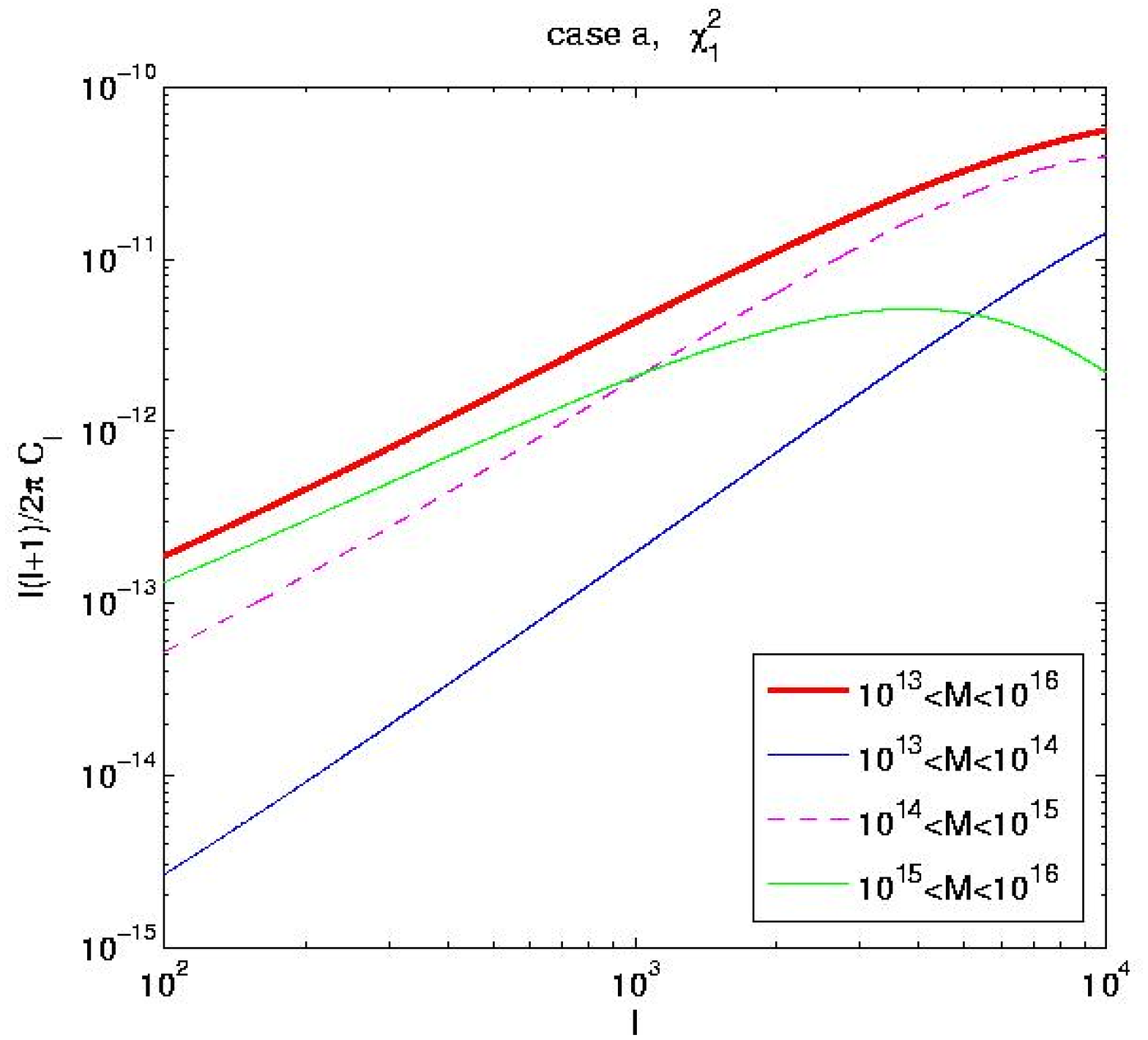}
\vspace{9pt}
\epsfxsize=2.5in
\includegraphics[width=6.5cm,height=6cm]{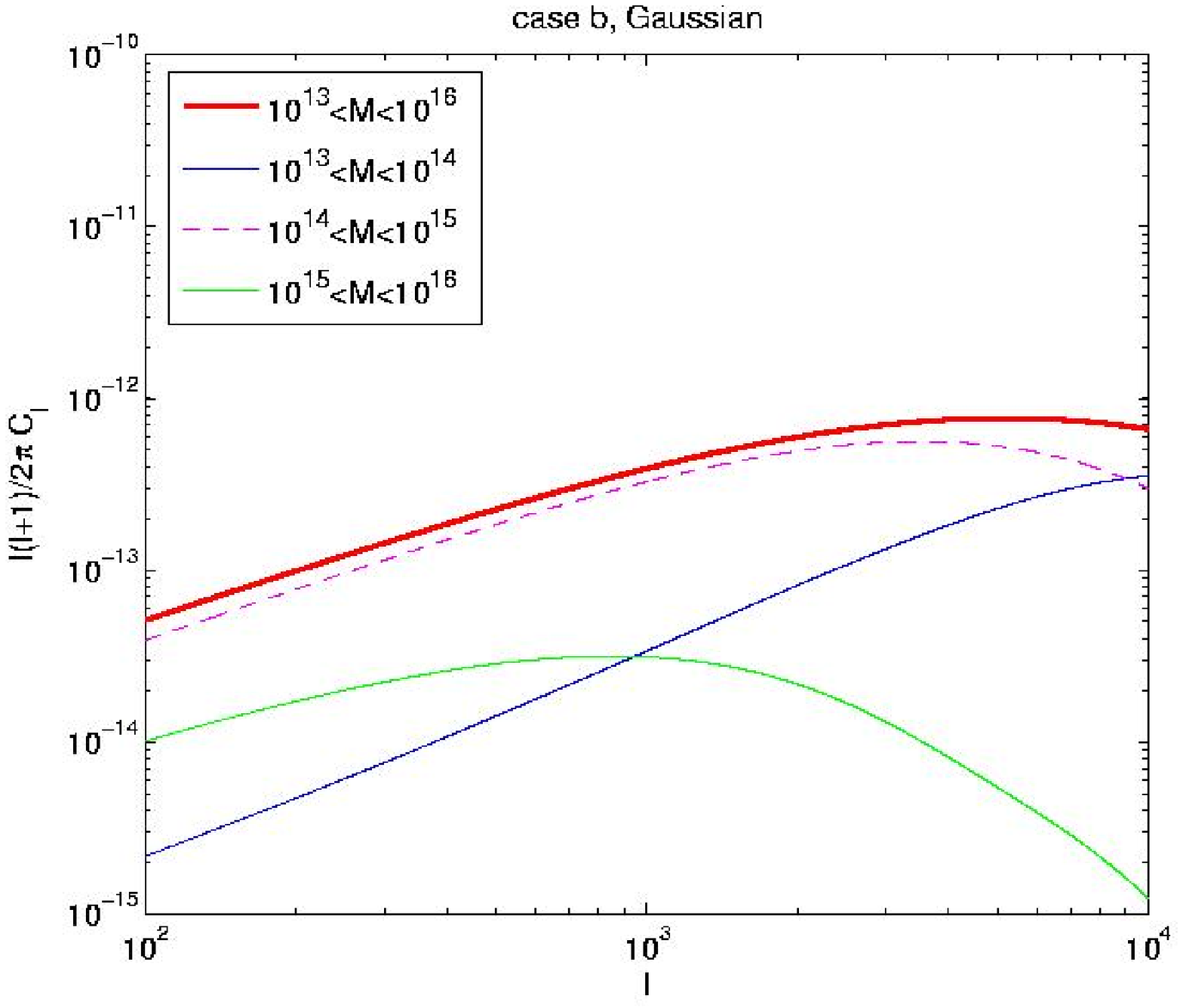}
\hspace{0.1in}
\includegraphics[width=6.5cm,height=6cm]{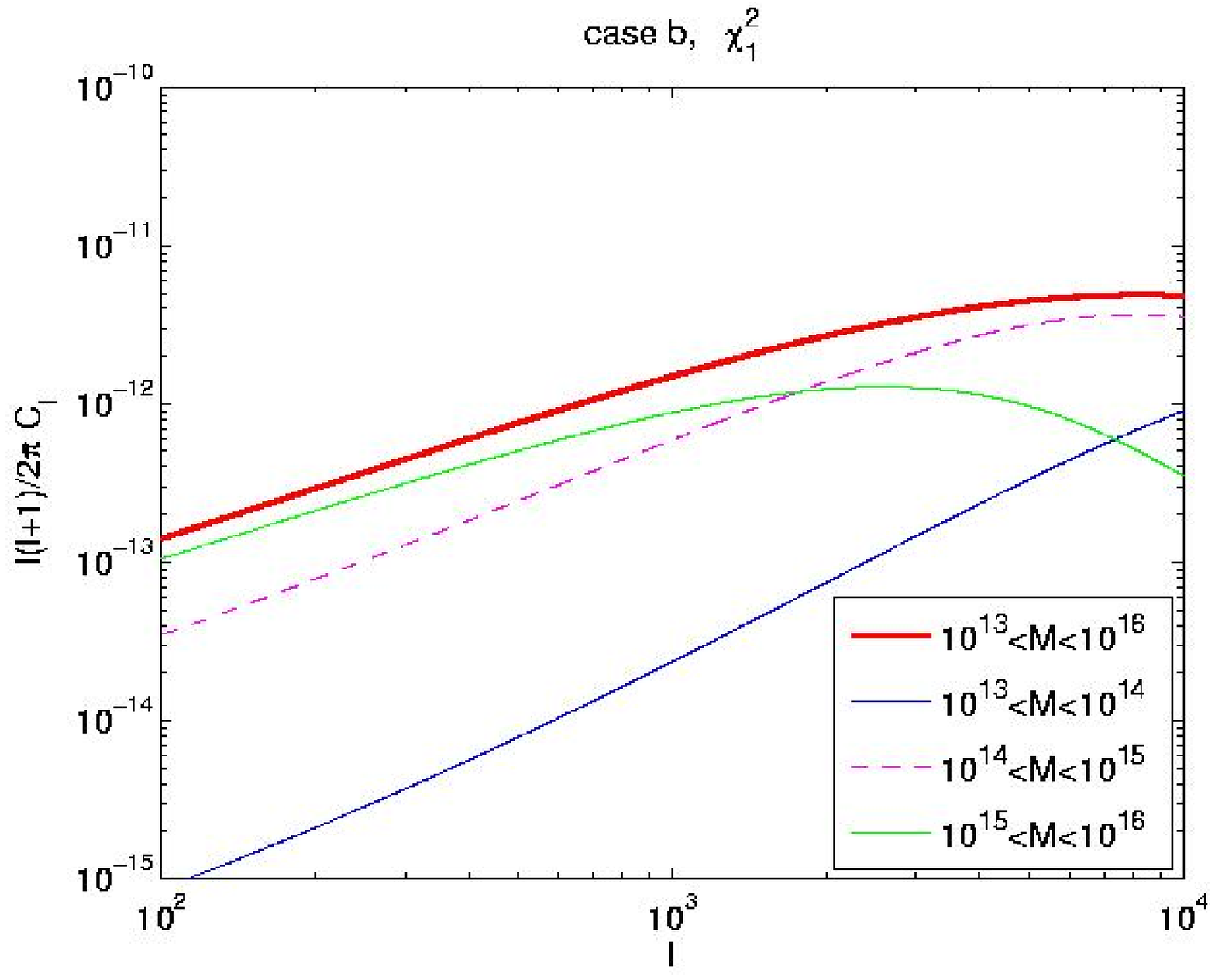}
\caption{As in the previous figure, but here levels of the power spectra 
are calculated for four mass intervals spanning the range 
$10^{13}h^{-1}M_{\odot}<M<10^{16}h^{-1}M_{\odot}$.}
\label{fig:clm}
\end{figure*}

Power spectra calculated over the redshift interval $[0,6]$ for cases 
(a) and (b) are plotted in Fig.~\ref{fig:clz} and Fig.~\ref{fig:clm}. 
In these figures the upper and lower panels correspond to cases (a) and 
(b), and the left and right panels are for the Gaussian and $\chi^{2}_1$ 
models, respectively. The figures show the distribution of the power spectrum 
with redshift (Fig.~\ref{fig:clz}) in $\Delta z=1$ and in (logarithmic) 
mass intervals of $\log{\Delta M}=1$ (Fig.~\ref{fig:clm}), respectively. 
The maximum power levels attained in case (a) are $\sim 3\cdot 10^{-12}$ 
and $\sim 6\cdot 10^{-11}$ for the Gaussian and non-Gaussian models, 
respectively. Both spectra peak at multipoles higher than $\ell>10,000$; 
in the Gaussian model this is due to the high gas densities of distant 
clusters, whereas in the non-Gaussian model this behavior is attributed 
to the combination of high gas densities and the long high-mass tail. 

Distant clusters are cooler in case (b) than in case (a), due to the 
redshift independence of their temperatures, so that power levels are lower 
($\sim 8\cdot 10^{-13}$ and $5\cdot 10^{-12}$) in the Gaussian and 
$\chi^{2}$ models, respectively. In the Gaussian model the power peaks at 
$\ell\sim 6000$, reflecting the lower contribution from cooler, more 
distant clusters. In the non-Gaussian model, a relatively high abundance 
of massive clusters at high redshifts leads to sustained high power levels 
on these scales. Note that irrespective of whether or not the temperature 
changes with redshift, it still increases with increasing mass, so even 
when $\psi=0$ a more abundant population of hot clusters is present in 
the $\chi^{2}$ model. In the $\chi^{2}$ model the contribution of 
relatively high $z$ clusters to the power is quite evident; in case (a) 
the contribution of clusters lying in the redshift range $0<z<1$ peaks 
at $\ell\sim 3000$, but the total power continues to rise due mostly 
to clusters in the range $1<z<2$. The same behavior, though less 
pronounced, is evident in case (b).
\begin{figure*}
\centering
\includegraphics[width=6.5cm,height=6cm]{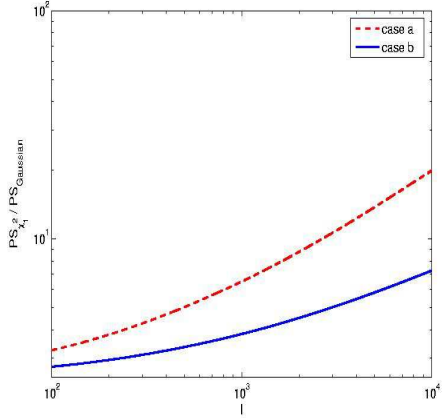}
\caption{A comparison of S-Z power spectrum levels in the $\chi^2_1$ 
and Gaussian models for cases (a) and (b).}
\label{fig:clcomp}
\end{figure*}

The distribution of S-Z power in four mass intervals is plotted in 
Fig.~\ref{fig:clm}. Of course, the highest mass interval 
($10^{15}h^{-1}M_{\odot}<M<10^{16}h^{-1}M_{\odot}$ is most 
significant in distinguishing between the signatures of Gaussian and 
non-Gaussian models. Clusters with masses in this range contribute negligibly 
at all multipoles in the Gaussian model (for which most of the power is 
produced by clusters in the mass range 
$10^{14}h^{-1}M_{\odot}<M<10^{15}h^{-1}M_{\odot}$), whereas they 
dominate the total power up to $\ell\sim 1000$ and $\ell\sim 2000$ 
in cases (a) and (b), respectively. At higher multipoles the 
$10^{14}h^{-1}M_{\odot}<M<10^{15}h^{-1}M_{\odot}$ range dominates. In 
the Gaussian model the contribution of low-mass clusters in the range 
$10^{13}h^{-1}M_{\odot}<M<10^{14}h^{-1}M_{\odot}$ is dominant only 
at the highest multipoles. To contrast the predictions of the two models, the 
ratio of total level of the S-Z power in the $\chi^2_1$ model to that in the 
Gaussian model for cases (a) and (b) is shown in Fig.~\ref{fig:clcomp}. For 
$\ell =10^3 - 10^4$, this ratio increases from $7$ to $20$ in case (a), 
and from $\sim 4$ to $\sim 7$ in case (b).
\begin{figure*}
\centering
\includegraphics[width=6.5cm,height=6cm]{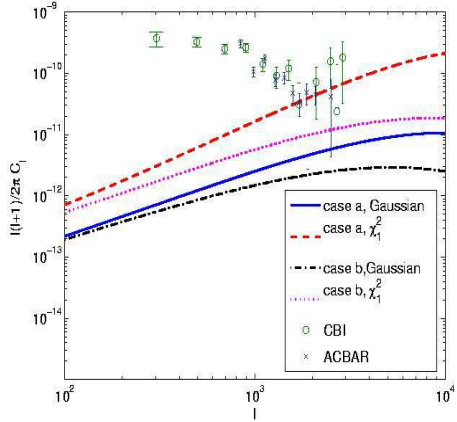}
\caption{Predicted S-Z power spectra at 31 GHz together with \textit{CBI} 
and \textit{ACBAR} datapoints.}
\label{fig:cbiacb}
\end{figure*}

Perhaps the strongest motivation for a non-Gaussian PDF is the apparent 
difficulty in interpreting the power measured at high $\ell$ by 
\textit{CBI} (Mason et al. 2003) and \textit{ACBAR} (Kuo et al. 2004) 
as (mostly) S-Z induced anisotropy in a Gaussian model. The predicted power 
spectra in the Gaussian and $\chi^2_1$ models are shown in 
Fig.~(\ref{fig:cbiacb}) together with the \textit{CBI} and 
\textit{ACBAR} measurements. The results of the calculations were 
appropriately adjusted to 31 GHz, the observation frequency of 
\textit{CBI}. It seems difficult to reconcile the Gaussian results with 
the data, while the predicted S-Z power in case (a) of the $\chi^2_1$ model 
matches reasonably well the observed power `excess' (over the extrapolated 
power measured by WMAP) at $\ell\geq 2000$ (see also Mathis et al. 2004). 
\begin{figure*}[b]
\centering
\vspace{9pt}
\epsfxsize=2.5in
\includegraphics[width=6.5cm,height=6cm]{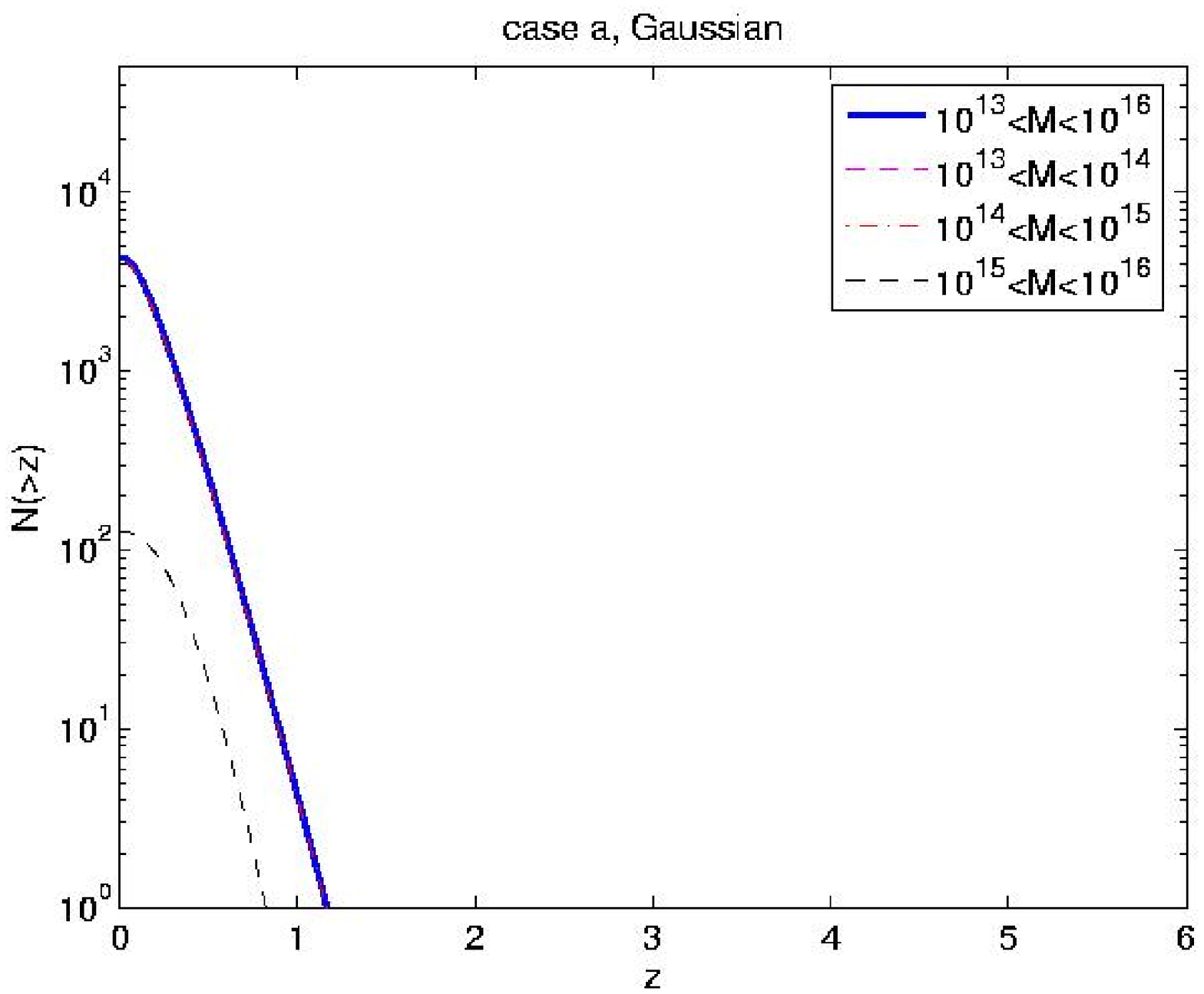}
\hspace{0.1in}
\includegraphics[width=6.5cm,height=6cm]{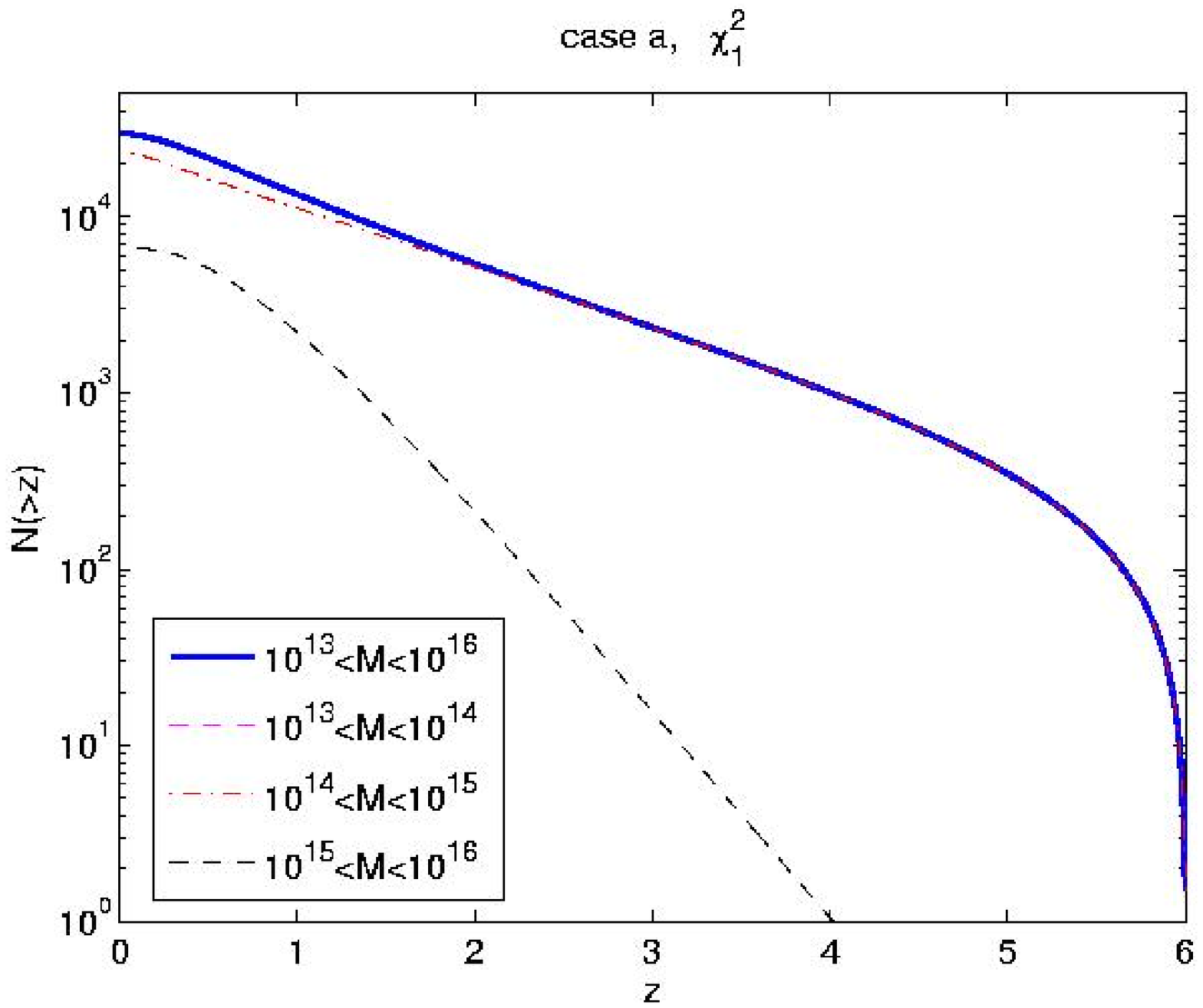}
\vspace{9pt}
\epsfxsize=2.5in
\includegraphics[width=6.5cm,height=6cm]{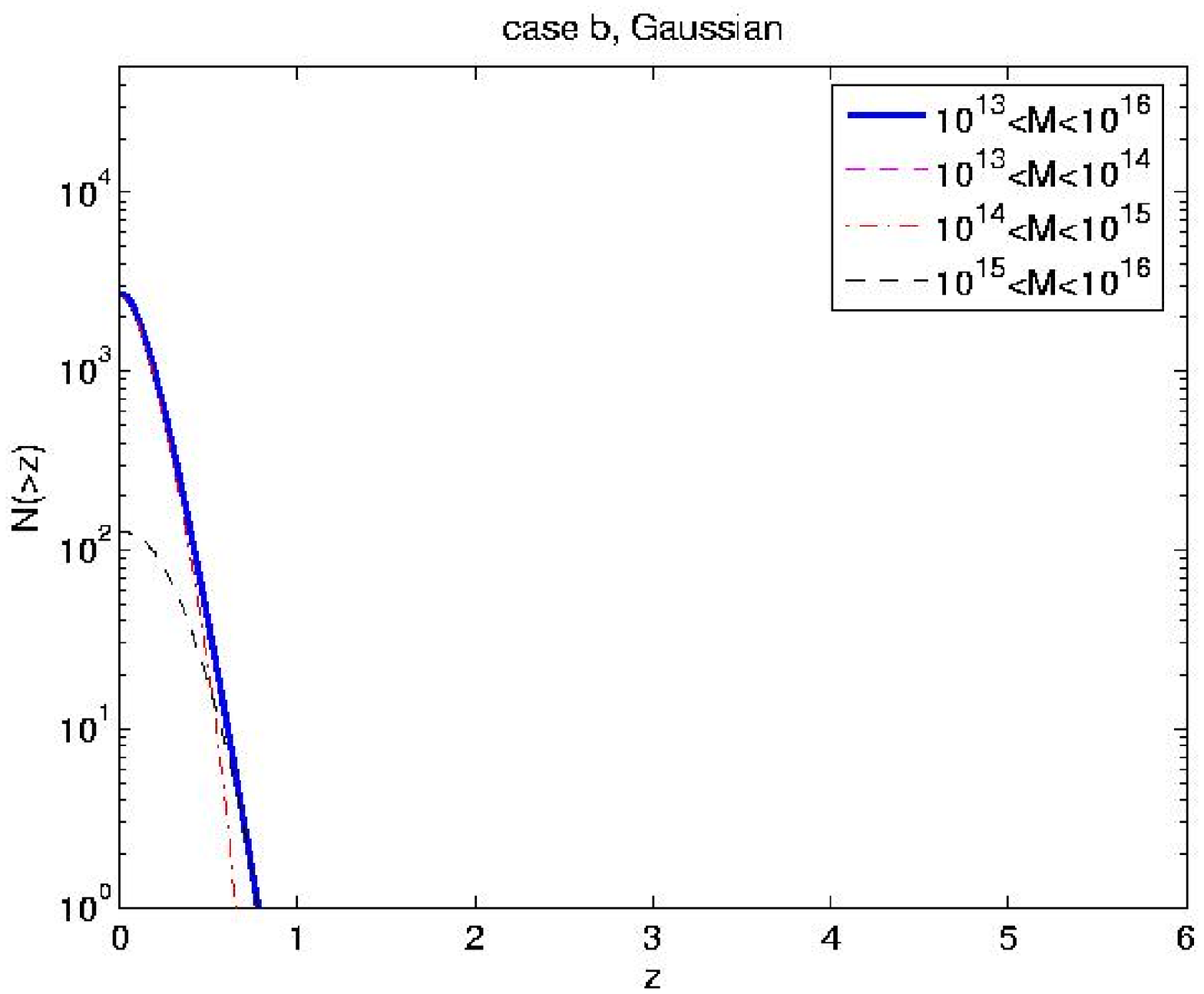}
\hspace{0.1in}
\includegraphics[width=6.5cm,height=6cm]{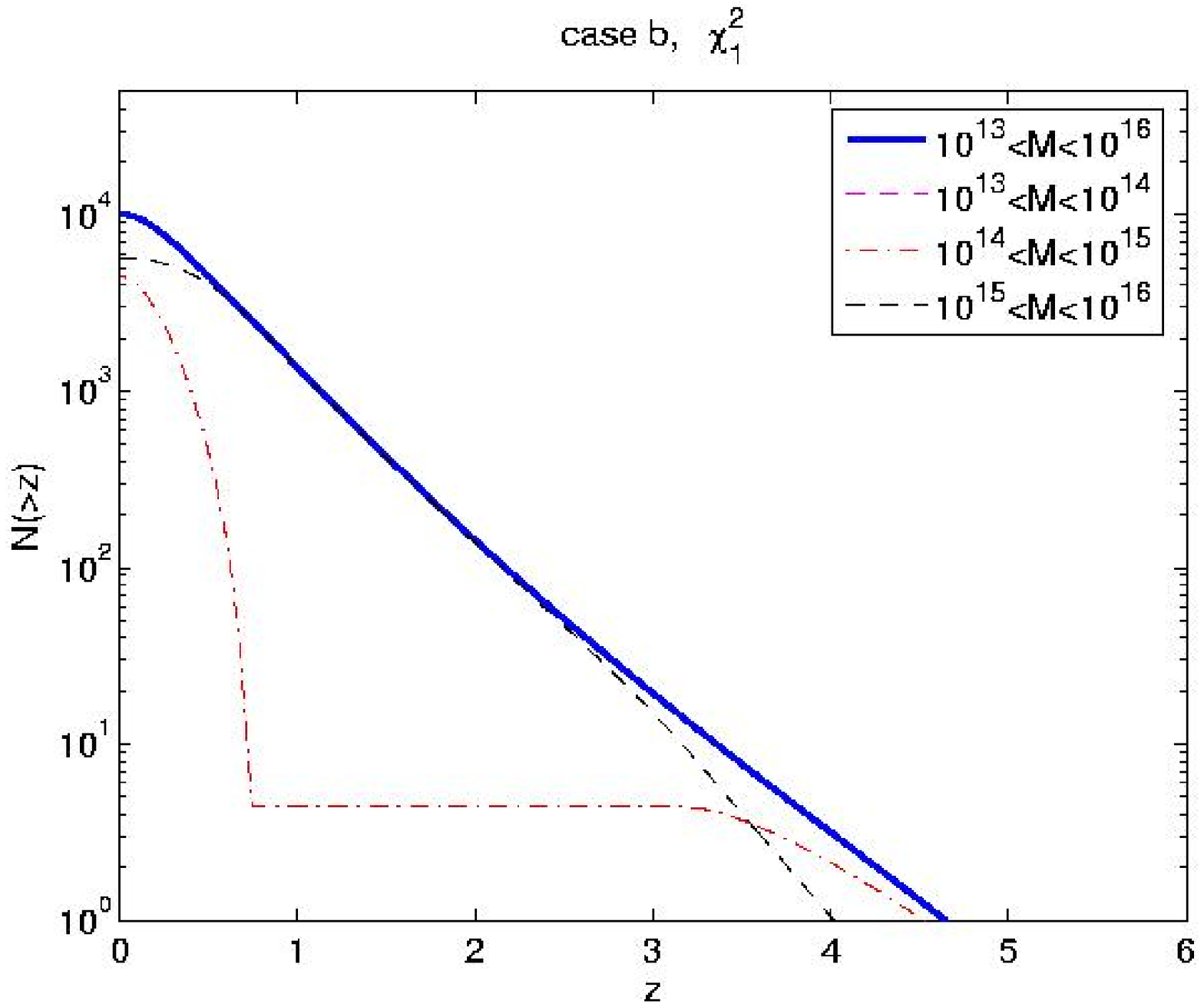}
\caption{Cumulative S-Z counts as function of redshift in four mass 
intervals. (The contribution from the lowest mass interval,
$10^{13}<M<10^{14} \, M_{\odot}$ , vanishes in all the cases 
considered here.}
\label{fig:nctz}
\end{figure*}

A more direct manifestation of the enhanced high-mass non-Gaussian PDF 
is apparent in the higher (cumulative) counts in this model, and 
a redshift distribution that is different than in the Gaussian model. 
In Fig.~\ref{fig:nctz} we show the redshift distribution of cumulative 
S-Z number counts. The plots are arranged by the specific model (as in 
Fig.~\ref{fig:clz} and Fig.~\ref{fig:clm}). Note that in all models 
considered here only clusters with masses higher than 
$10^{14}h^{-1}M_{\odot}$ generate fluxes that exceed the adopted 
detection limit of $30\,mJy$. 
\begin{figure*}
\centering
\vspace{9pt}
\epsfxsize=2.5in
\includegraphics[width=6.5cm,height=6cm]{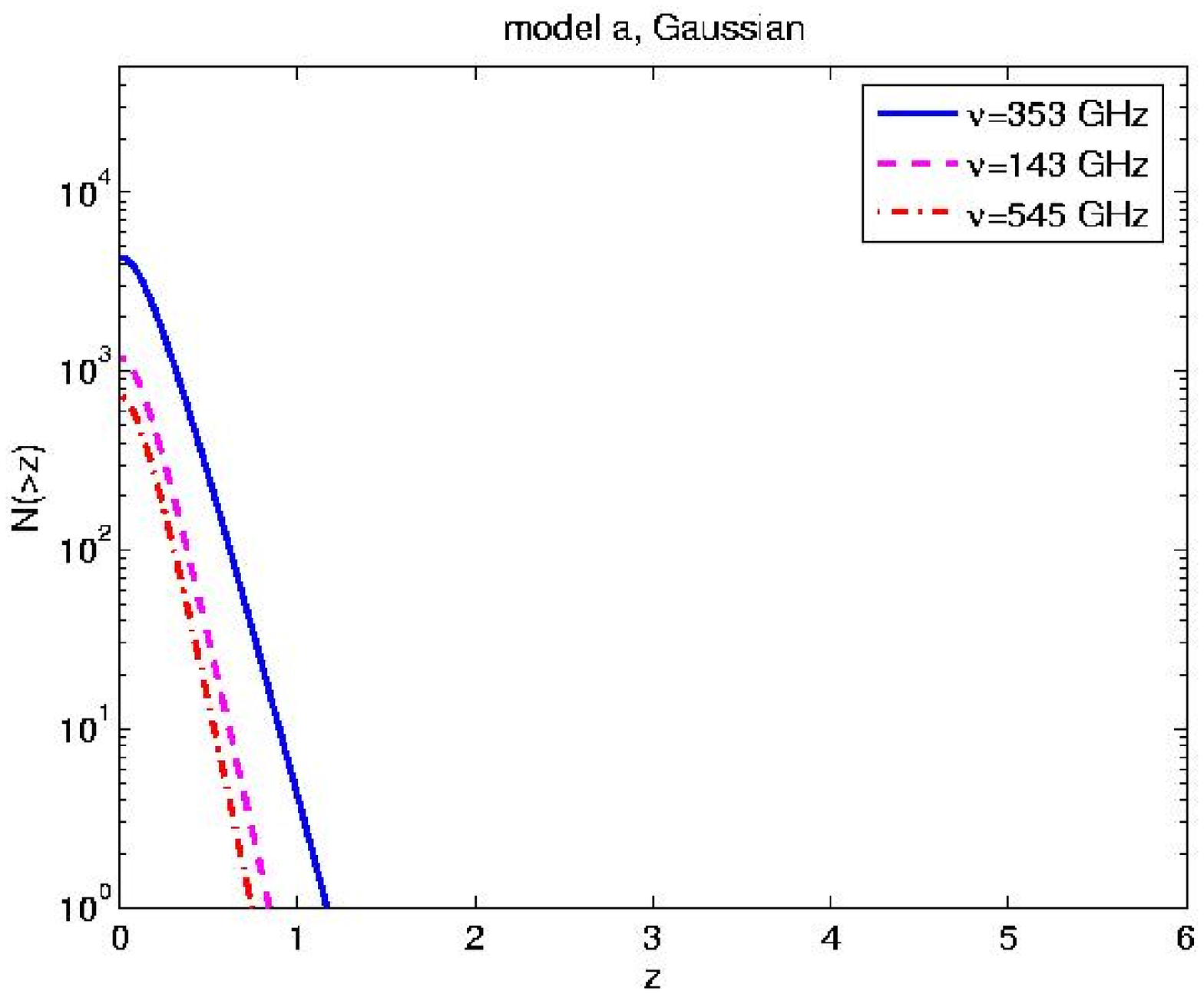}
\hspace{0.1in}
\includegraphics[width=6.5cm,height=6cm]{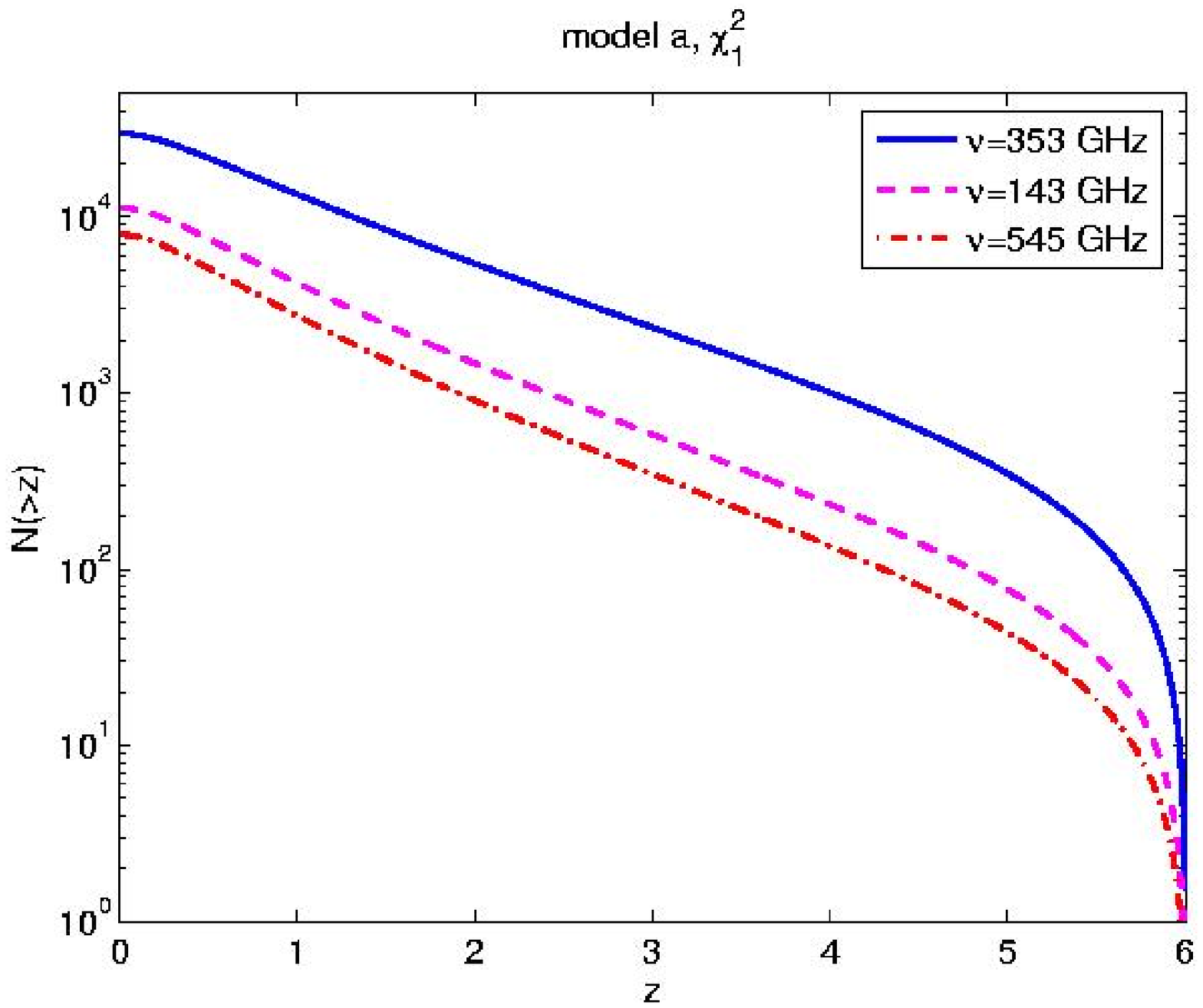}
\vspace{9pt}
\epsfxsize=2.5in
\includegraphics[width=6.5cm,height=6cm]{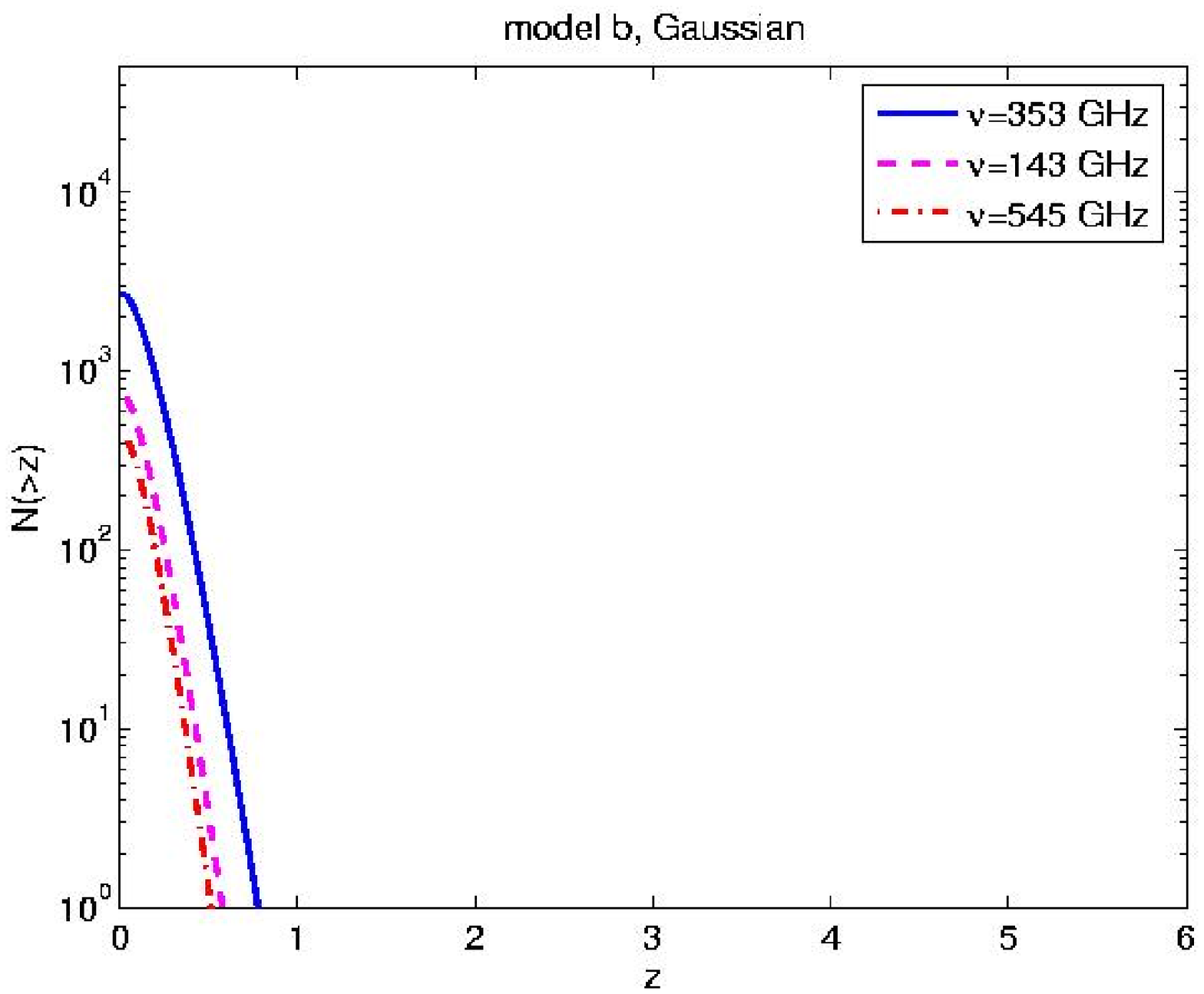}
\hspace{0.1in}
\includegraphics[width=6.5cm,height=6cm]{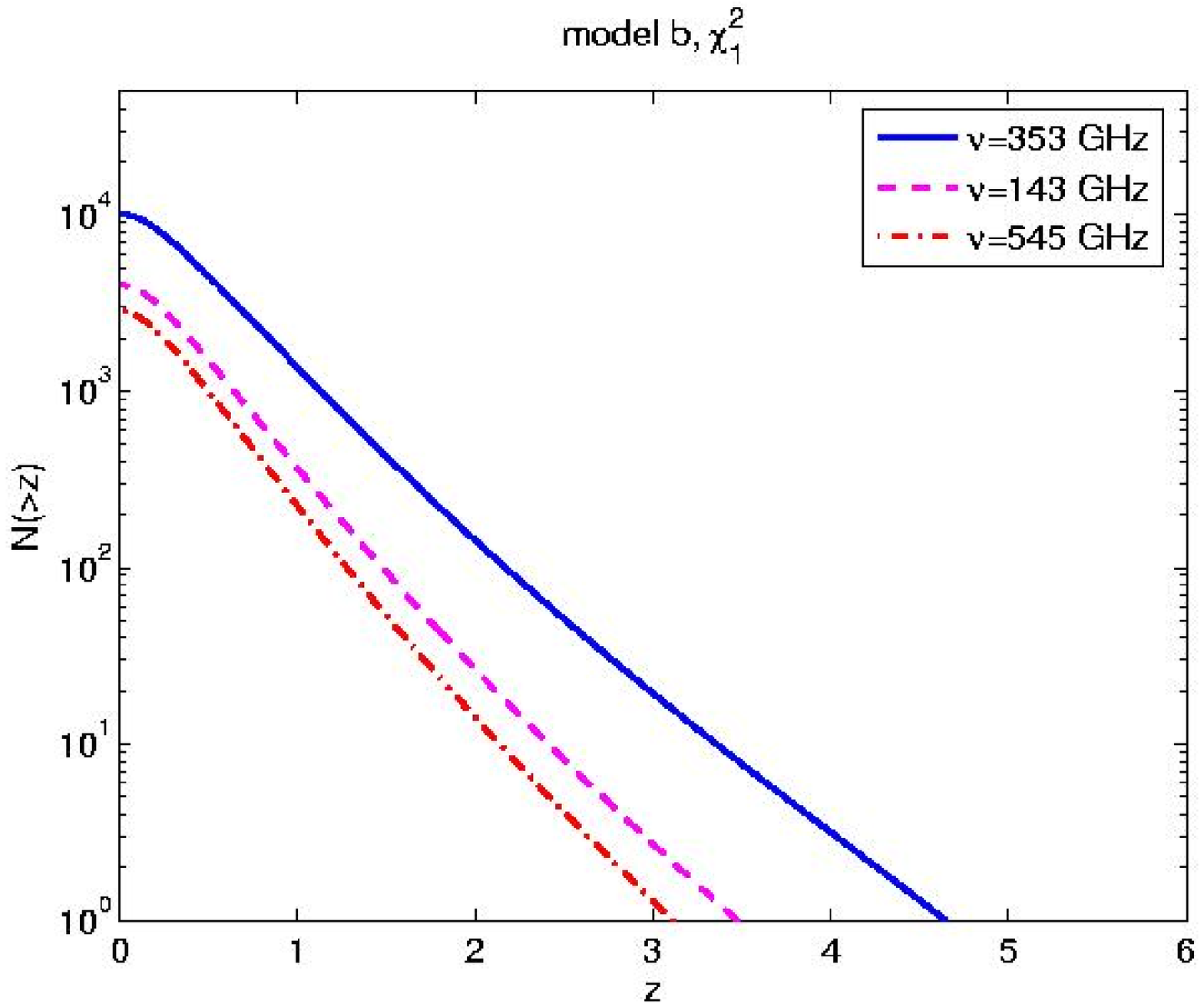}
\caption{Cumulative S-Z counts as a function of redshift. Number 
counts are shown for the three specified frequency channels of the 
HFI instrument on the \textit{Planck} satellite.} 
\label{fig:nctz1}
\end{figure*}

Total counts in the Gaussian model are $\sim 4000$ and $\sim 3000$ in 
cases (a) and (b), respectively, whereas these are $\sim 30,000$ and 
$\sim 10,000$ in the corresponding cases of the $\chi^2_1$ model. In 
the Gaussian model the largest contribution to the number counts 
(at $z=0$) comes from clusters in the mass range 
$10^{14}h^{-1}M_{\odot}<M<10^{15}h^{-1}M_{\odot}$, whose numbers 
are at least an order of magnitude higher than those in the mass range 
$10^{15}h^{-1}M_{\odot}<M<10^{16}h^{-1}M_{\odot}$. In case (a) of 
the $\chi^2_1$ model, total counts are dominated by clusters in the mass 
range $10^{14}h^{-1}M_{\odot}<M<10^{15}h^{-1}M_{\odot}$, but 
differences in the relative contribution of higher mass clusters are not 
as pronounced as in the Gaussian case. On the other hand, in case (b) in 
this model the total counts are dominated by high-mass clusters at 
$0\leq z\leq 3.5$. The cumulative counts at the lowest redshifts are 
slightly higher (by a factor $\sim 1.3$) in the high-mass range than 
the corresponding contribution from clusters in the mass range 
$10^{14}h^{-1}M_{\odot}<M<10^{15}h^{-1}M_{\odot}$. Cluster counts 
are also more broadly distributed in redshift space in the non-Gaussian 
model due to their higher abundance at high redshifts. 

In Fig.~\ref{fig:nctz1} we plot the cumulative number counts for 
two more of \textit{Planck} frequency channels, $\nu=143$ GHz and 
$\nu=545$ GHz (in addition to $\nu=353$ GHz). In the non-relativistic 
electron velocity limit the (thermal component of the) S-Z spectral 
function is $g(x)$. At these frequencies, $g(x)=-4, 3.2, 6.7$, 
respectively, so our results for $\nu=353$ GHz can 
be scaled accordingly. Adopting the same value of the limiting flux 
($30$ mJy), we find that values of the ratios of number counts at 
$\nu=353$ GHz and $\nu=545$ GHz are larger in the Gaussian than in 
the $\chi^2_1$ model. This can clearly be attributed to the significantly 
higher population of clusters with flux exceeding the limit in the 
$\chi^2_1$ model. Note that an accurate calculation of the S-Z 
intensity change necessitates a relativistic calculation (Rephaeli 
1995) which results in a more cumbersome expression for the 
temperature-dependent spectral function (for which there are 
a few analytic approximations; see, e.g., Shimon \& Rephaeli 2004, Itoh 
\& Nozawa 2004). The deviation from the \nrel calculation can amount 
to few tens of percent for typical temperatures in the range 5-10 keV. 
Since our main focus here is a comparison between predictions of the 
two density fields for quantities that are integrated over the cluster 
population (rather than an accurate description of the effect in a 
given cluster), it suffices to use the much simpler function $g(x)$. 

The impact of varying $\alpha$ and $r_c$ on the predicted power 
spectra and cluster number counts in the Gaussian and $\chi^{2}_1$ 
models was explored by SRS. Here we only note that the observed mean 
ranges of these parameters do not introduce appreciable uncertainty 
to blur the significant differences between the predictions of the 
two models. SRS have also investigated the consequences of the 
$\chi^{2}_1$ model on cluster formation times and the two-point 
correlation function of clusters.

In summary, the main objective of the work of SRS has been to compare 
predictions of S-Z observables for two mass functions which differ mainly 
at the high mass end. A $\chi^2_1$-distributed field is characterized 
by a longer tail of high density fluctuations with respect to the Gaussian 
model. This leads to the formation of high-mass clusters at higher 
redshifts. S-Z power spectra, number counts (and two-point angular 
correlation functions) yield significantly different results in the 
$\chi^2_1$ primordial density field as compared with the respective 
quantities in a Gaussian model.

\section{Polarization Patterns in a Simulated Cluster}

Unpolarized incident radiation is linearly polarized by Compton scattering 
when the radiation either has a quadrupole moment, or acquires it during 
the (first stage of the) scattering process. The degree of polarization due 
to scattering is proportional to the product of the quadrupole moment and the 
Compton optical depth, $\tau$. Thus, the largest polarization levels in 
clusters are expected to be at the few $\mu$K level; nonetheless, the 
detection of polarized CMB signals at this level is presumably feasible 
(Bowden \ea 2004). The prospects for measurements of polarized S-Z signals 
motivate investigation of their levels and patterns.  

Several polarization components due to scattering of the CMB in clusters were 
described in detail by Sunyaev \& Zeldovich (1980) and Sazonov \& Sunyaev 
(1999). In these treatments simple models for cluster morphology and IC gas 
spatial profile were assumed. Since polarization levels and patterns strongly 
depend on the (generally) non-uniform distributions of IC gas and total mass, 
on peculiar and internal velocities, and also on the gas temperature profile, 
a realistic characterization of the various cluster polarization components 
can only be based on the results of hydrodynamical simulations. This is 
particularly the case in clusters with a substantial degree of subclustering 
and merger activity.

We began a study of S-Z polarization in non-idealized clusters based on 
cosmological hydrodynamical simulations of the formation and evolution of 
clusters. Here we discuss first results for the polarization structure of the 
properties of the S-Z effect. In particular, we consider the two polarization 
components associated with the thermal and kinematic effects that are 
generated when the CMB is doubly scattered, i.e. the polarization is 
$\propto \tau^2$. In the first, the initial anisotropy arises from the 
thermal S-Z effect, whereas in the second the initial anisotropy is produced 
by scattering in a cluster with a velocity component transverse to the los, 
$v_t \equiv c\beta_{t}$. The spatial patterns of these components can be 
readily determined only when the morphology of the cluster and its IC gas are 
isotropic. Polarization patterns arising from scattering off thermal 
electrons are isotropic (Sazonov \& Sunyaev 1999) in a spherical cluster, 
while patterns of all kinematic polarization components are always 
anisotropic due to the asymmetry introduced by the direction of the cluster 
motion.

\subsection{Kinematic and Thermal Polarization Components}

Linear polarization and its orientation are determined by the two Stokes 
parameters
\begin{eqnarray}
Q&=&\frac{3\sigma_{T}}{16\pi}\int n_{e} dl \int \sin^{2}\alpha\cos
2\psi I(\alpha,\psi) d\Omega \nonumber\\
U&=&\frac{3\sigma_{T}}{16\pi}\int n_{e} dl\int \sin^{2}\alpha\sin
2\psi I(\alpha,\psi) d\Omega ,
\end{eqnarray} 
where $dl$ is a length element along the photon path, and $\alpha$ and 
$\psi$ define the relative directions of the incoming and outgoing photons. 
The average electric field describes the polarization plane whose orientation 
is determined by the angle 
\begin{eqnarray}
\varphi=\frac{1}{2}\tan^{-1}\frac{U}{Q}.
\end{eqnarray}
When the incident radiation is expanded in spherical harmonics
\begin{eqnarray}
I(\alpha,\psi)=\sum_{l,m}I_{lm}Y_{lm}(\alpha,\psi),
\end{eqnarray}
it can be seen (from the orthogonality conditions of the spherical harmonics) 
that only the quadrupole moment terms proportional to $I_{2,2}$ and 
$I_{2,-2}$ contribute to $Q$ and $U$.

The expressions for the $Q$ and $U$ in the equation (22) are given in terms 
of the relative directions of the ingoing and outgoing photons. Due to the 
obvious dependence of the angles and frequencies on the relative directions 
of the electron and photon motions, these expressions are transformed to a 
frame whose $Z$ axis coincides with the direction of the electron velocity, 
the axis with respect to which the incoming and outgoing photon directions 
are defined (in accord with the choice made by Chandrasekhar 1950). Since 
the Doppler effect depends only on the angle between the photon wave vector 
and the electron velocity, the calculation of the temperature anisotropy and 
polarization is easier if we assume that the incident radiation is isotropic 
in the CMB frame. With the polarized cross section written in terms of 
angles measured with respect to the electron velocity (after averaging over 
the azimuthal angles),
\begin{eqnarray}
\frac{d\sigma}{d\Omega}=\frac{3\sigma_{T}}{8}
\left(1-\mu_{0}^{2}\right)P_{2}(\mu'_{0})\, ,
\end{eqnarray}
where $P_{2}(\mu'_{0})$ is the second Legendre polynomial, the expression 
for the $Q$ parameter of the scattered radiation is 
\begin{eqnarray}
Q(\mu)=\frac{3}{8}\tau(1-\mu_{0}^{2})
\int_{-1}^{1}P_{2}(\mu'_{0})I(\mu'_{0})d\mu'_{0}\, .
\end{eqnarray}
Here, $\mu'_{0}=\cos\theta'_{0}$, $\mu_{0}=\cos\theta_{0}$, are the 
angle cosines between the electron velocity and the incoming and outgoing 
photons, respectively. 

The degree of polarization induced by the kinematic S-Z component was 
determined by Sunyaev \& Zeldovich (1980) in the simple case of uniform 
gas density,
\begin{eqnarray}
Q=\frac{1}{40}\frac{x e^{x}}{e^{x}-1} \tau^{2} \beta_{t} ,
\end{eqnarray}
where (as defined in the previous section) $x=h\nu/kT$. A more complete 
calculation of this and the other polarization components 
was formulated by Sazonov \& Sunyaev (1999). Viewed along a direction 
$\hat{n}=(\theta,\phi)$, the temperature anisotropy at a point $(X,Y,Z)$, 
$\Delta T(X,Y,Z,\theta,\phi)$ generates polarization upon second 
scatterings. The Stokes parameters are calculated from equation (22), 
\begin{eqnarray}
Q(X,Y)&=&\frac{3\sigma_{T}}{16\pi I_{0}}\int dZ n_{e}(X,Y,Z)\nonumber\\
&&\times\int d\Omega\sin^{2}(\theta)\cos(2\phi)\Delta I(X,Y,Z,\theta,\phi)\nonumber\\
U(X,Y)&=&\frac{3\sigma_{T}}{16\pi I_{0}}\int dZ n_{e}(X,Y,Z)\nonumber\\
&&\times\int d\Omega\sin^{2}(\theta)\sin(2\phi)\Delta I(X,Y,Z,\theta,\phi)
\end{eqnarray}
where now the $Z$ direction is chosen along the los. The intensity change 
resulting from first scatterings is 
\begin{eqnarray}
& &\Delta I(X,Y,Z,\theta,\phi)/I_{0}=\frac{\sigma_{T}x e^{x}}{e^{x}-1},
\int d\vec{l}(X',Y',Z',\theta,\phi)\nonumber\\
& &\times n_{e}(X',Y',Z',\theta,\phi)\hat{n}\cdot 
\beta(X',Y',Z')
\end{eqnarray}
and the optical depth through the point $(X,Y,Z)$ in the direction 
$(\theta,\phi)$ is
\begin{eqnarray}
\tau(X,Y,Z,\theta,\phi)=\sigma_{T}\int n_{e}(X',Y',Z')d\vec{l}
(X',Y',Z',\theta,\phi).
\end{eqnarray}
$Q(X,Y)$ and $U(X,Y)$ fully describe the 2-D polarization field.

The anisotropy introduced by single scatterings off thermal electrons 
is the usual thermal component of the S-Z effect with intensity change 
$\Delta I_{t}$. Using the analytic approximation to the exact 
relativistic calculation for $\Delta I_{t}$ that was obtained by Shimon 
\& Rephaeli (2004), the polarization can be calculated as described above 
\begin{eqnarray}
\Delta I_{t}(X,Y,Z,\theta,\phi)
=\frac{\sigma_{T}I_{0}x^{3}}{e^{x}-1}
\sum_{i=1}^{5}f_{i}(x)\int n_e({\bf r})\Theta({\bf r})^{i}dl, 
\end{eqnarray}
where $\Theta=kT_{e}/mc^2$ and $T_{e}$ is the electron temperature. 
In Equation (31) the integration is along the photon path prior to the 
second scattering, $f_{i}(x)$ are spectral functions of $x$ that are 
specified by Shimon \& Rephaeli (2004).

\subsection{Clster Simulations}

We investigated the polarization properties induced by scattering 
in a cluster with (a total) mass of $1.4\times 10^{15}\, M_{\odot}$ 
simulated with Enzo, an Eulerian adaptive mesh cosmology code (O'Shea 
\ea 2004). The code solves dark matter N-body dynamics with the particle 
mesh technique, and Euler's hydrodynamic equations with the piecewise 
parabolic method algorithm modified for cluster simulations by Bryan 
et al. (1995). The simulation was initialized with dark matter in a 
volume of space 256h$^{-1}$ Mpc on a side at $z=30$ (when density 
fluctuations are still linear on this scale), using a 256$^3$ root 
grid with 256$^3$ dark matter particles (with mass 
$M_{DM} = 1.19\times 10^{11}~M_{\odot}$). The standard 
$\Lambda$CDM cosmological model was used with $\Omega_{tot}=1$, 
$\Omega_{\Lambda}=0.7$, $\Omega_{M}=0.3$, $H_{0}=70$ km/sec/Mpc, 
$\sigma_{8}=0.9$, and $n=1$. Adaptive mesh refinement was turned on 
with an additional four levels of refinement (doubling the spatial 
resolution at each level) and the simulation was run to $z=0$. At this 
point the simulation was stopped and the Hop halo-finding algorithm 
(Eisenstein \& Hut 1998) was used to find the most massive dark matter 
halo in the simulation. This is our simulated 
$1.4 \times 10^{15}~M_{\odot}$ cluster. 

The simulation was then re-initialized at $z=30$ with both dark matter 
and baryons ($\Omega_{b} = 0.04,\, \Omega_{CDM} = 0.26$) with a 
128$^3$ top grid and 2 static nested grids covering the volume where the 
cluster forms, giving an effective root grid resolution of 512$^3$ cells 
($0.5h^{-1}$ Mpc comoving). With adaptive mesh refinement a maximum 
spatial resolution of 15.625 h$^{-1}$~kpc (comoving) was attained. 
This simulation was then evolved to $z=0.06$, following the evolution 
of the dark matter and using adiabatic gas dynamics. The cluster was then 
relocated with the Hop algorithm at $z=0.06$, $0.25$, $0.5$ and $1.0$, 
and data cubes with 256$^3$ cells containing $\rho_{DM}$, $\rho_{gas}$, 
$T_{e}$ and the three baryon velocity components were extracted at these 
redshifts. The total cube length is $4h^{-1}$ Mpc, corresponding to a 
spatial resolution of $15.625h^{-1}$ kpc (comoving). The results discussed 
here are based on the simulated cluster at $z=0.5$. The baryon density and 
velocity field (with velocity dispersion of $\approx 1000$ km/sec) are 
shown in Figure 7, and the cluster projected temperature and optical depth 
are mapped in Figure 8.

\subsection{Polarization Maps}

Specific calculations of the above double polarization components 
necessitate ray tracing of doubly scattered photons. In Cartesian 
coordinates, equations (28) \& (31) can be written as 
\begin{eqnarray}
Q(X,Y)&=&\frac{3\sigma_{T}^{2}}{16\pi}
\int\int dZd^{3}{\bf r'}n_{e}({\bf r})n_{e}({\bf r'})\nonumber\\
&&\times\frac{(X-X')^{2}-(Y-Y')^{2}}{|{\bf r'}-{\bf
    r}|^{4}}\sum_{i=1}^{5}f_{i}(x)\Theta({\bf r'})^{i}\nonumber\\
U(X,Y)&=&\frac{3\sigma_{T}^{2}}{8\pi}
\int\int dZd^{3}{\bf r'}n_{e}({\bf r})n_{e}({\bf r'})\nonumber\\
&&\times\frac{(X-X')(Y-Y')}{|{\bf r'}-{\bf r}|^{4}}
\sum_{i=1}^{5}f_{i}(x)\Theta({\bf r'})^{i}
\end{eqnarray}
where the first integration (over $Z$) is along the los. 
To carry out the computationally intensive 4D integrals over the simulation 
$256^{3}$ data cells, we write the last equation in the form 
\begin{eqnarray}
P_{+,\times}(X,Y)=\sigma_{T}^{2}\int dZ n_{e}({\bf r})h_{+,\times}({\bf r})
\end{eqnarray}
where $P_{+}$ and $P_{\times}$ are the Stokes $Q$ \& $U$ parameters,
respectively, and $h_{+,\times}=f_{+,\times}\star g$ is a convolution 
of the functions $f_{+,\times}$ with $g$, where
\begin{eqnarray}
f_{+}&=&\frac{X^{2}-Y^{2}}{r^{4}}, \, f_{\times}=\frac{2XY}{r^{4}},\, 
g=\sum_{i}n_{e}({\bf r})f_{i}(x)\Theta({\bf r})^{i}.
\end{eqnarray}
The 3D FFT, which can be handled considerably more efficiently, can now be 
calculated using the latter expressions for the core region of the simulation.

\begin{figure*}
\centering
\includegraphics[width=3.5in]{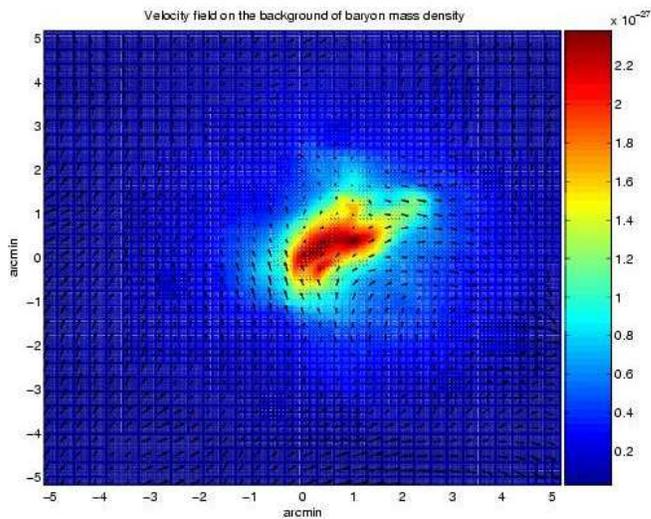}
\caption{The velocity field is shown on the background of the baryon mass 
density; arrow length is linearly proportional to velocity magnitude). 
Color density scale is in units of $g/cm^{3}$.}
\end{figure*}

\begin{figure*}
\centering
\vspace{9pt}
\epsfxsize=2.5in
\includegraphics[width=6cm,height=5.5cm]{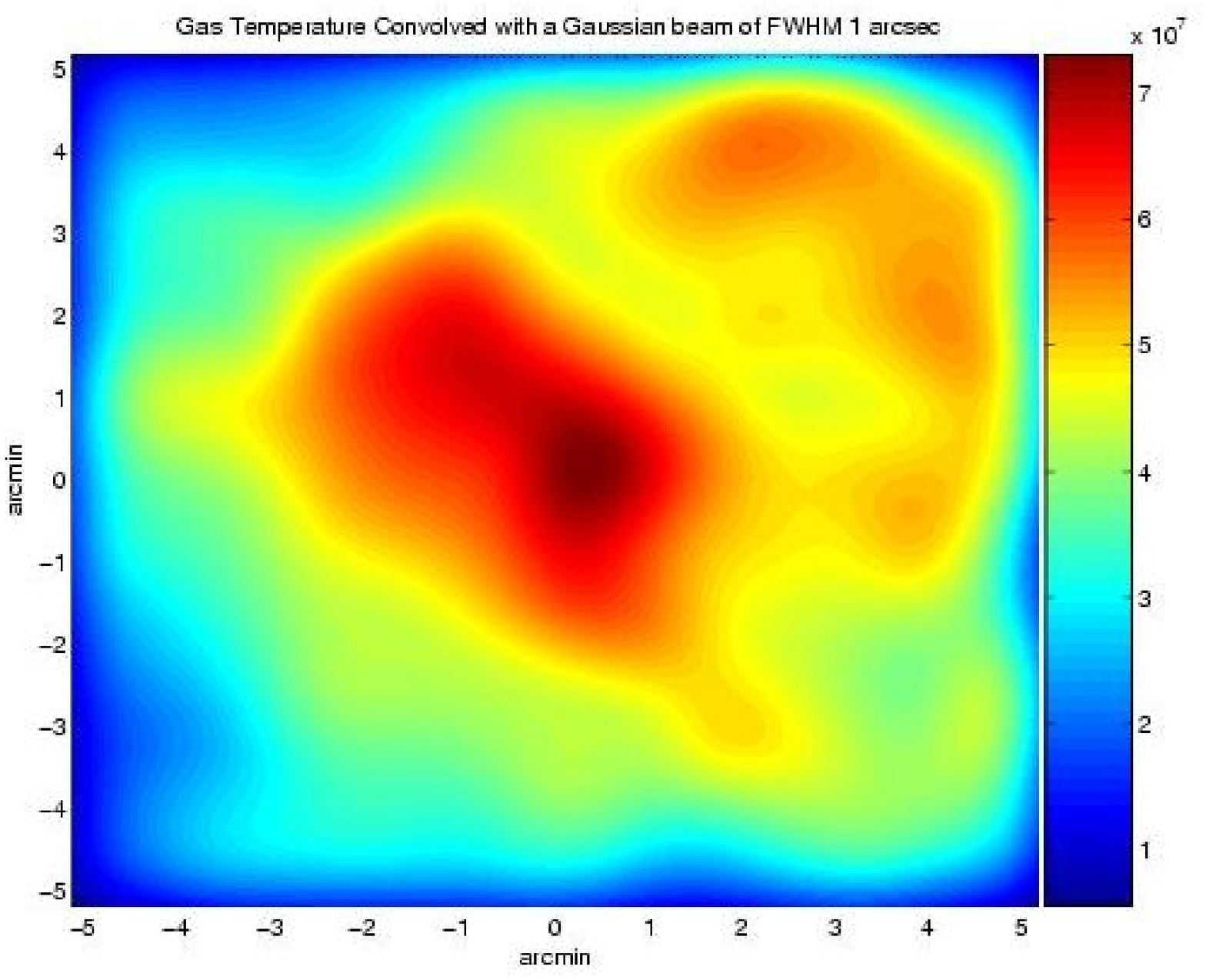}
\hspace{0.1in}
\includegraphics[width=6cm,height=5.5cm]{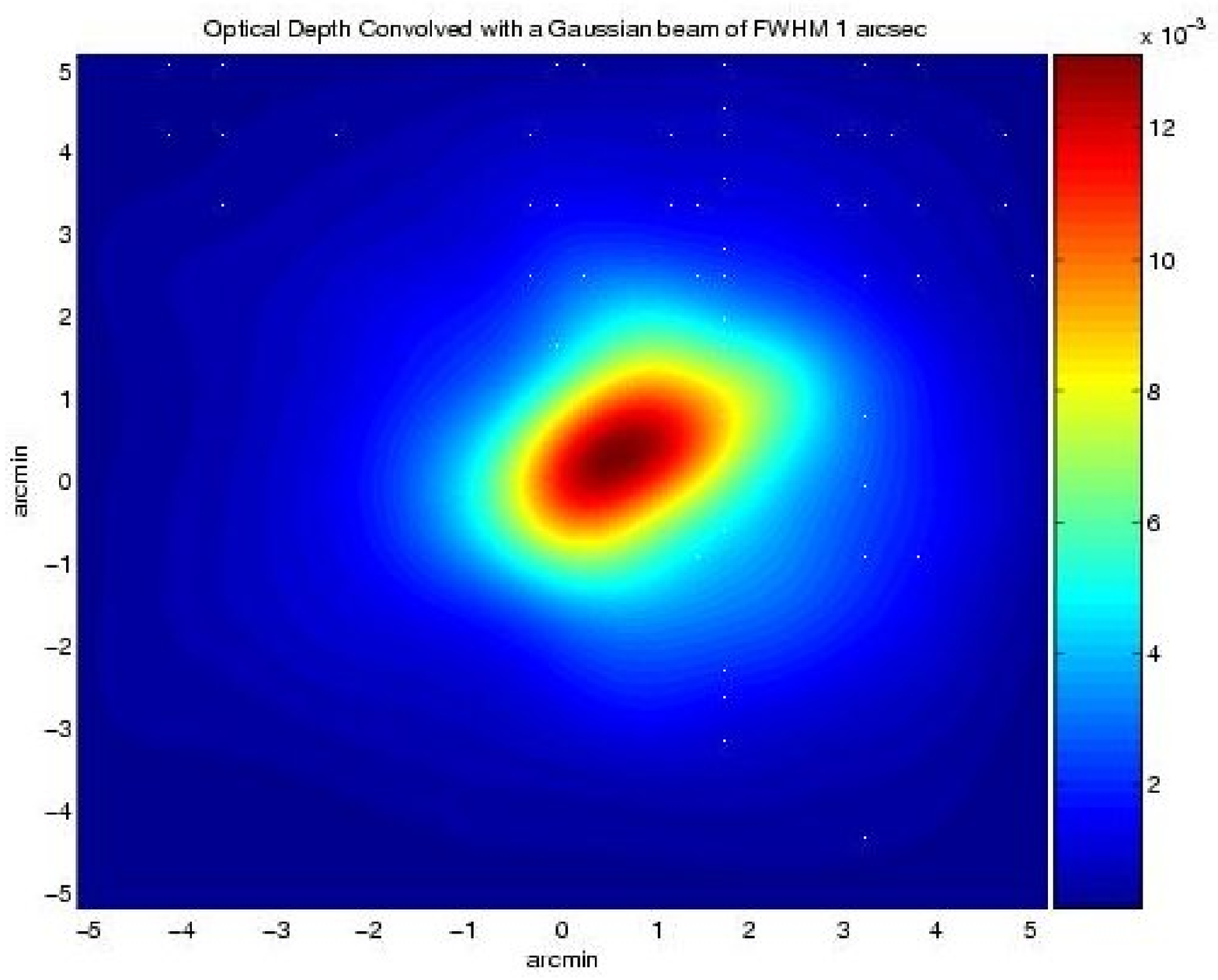}
\caption{The cluster temperature (color scale in K) and optical depth 
maps are shown in the left and right figures, respectively.}
\end{figure*}

Polarization maps on the sky plane are shown in Figures 9 \& 10, with 
polarization levels specified in terms of the equivalent temperature. 
Unlike the case of spherical (Sazonov \& Sunyaev 1999) or quasi-spherical 
(Lavaux et al. 2004) clusters, we obtain a more intricate polarization 
patterns and the quadrupole structure, which is absent in the spherically 
symmetric case, is evident. This pattern reflects the non-uniform gas 
distribution and an appreciable degree of non-sphericity of the cluster, 
which substantially affect primarily the double scattering polarization 
signals, essentially due to the fact that the second scattering is more 
likely to occur far from the cluster center (see also Sazonov \& Sunyaev 
1999, Lavaux \ea 2004). The level of polarization at the 545 GHz Planck 
frequency band, with a 4.5' FWHM beam, can reach $\sim 80$ nK, but 
when convolved with a narrow beam 1' FWHM profile, its peak value 
can be as high as $\sim 0.7\mu K$. This level definitely grazes the 
threshold of next generation experiments. It should be noted that we 
have considered a rich cluster with maximal optical depth of 
$\tau_{0}\approx 0.01$, but since this component is proportional 
to $\sim\tau^{2}$, typical values could be a factor $\sim 4$ lower 
or higher than the specific values quoted here. Also, this component 
was calculated at 545 GHz; lower levels of polarization are expected 
at lower frequencies.

Equations (28)-(30) for the double scattering kinematic component 
($\propto \tau^{2}\beta$) can be written in the same form as 
equation (33), but with
\begin{eqnarray}
h_{+,\times}=f_{+,\times}\star g\equiv\sum_{i=1}^{3}f_{i_{+,\times}}\star g_{i}
\end{eqnarray}
where
\begin{eqnarray}
f_{i+}=\frac{X_{i}(X^{2}-Y^{2})}{r^{5}} ,\,\,\,f_{i\times}=
\frac{2X_{i}XY}{r^{5}},\,\,\, g_{i}=n({\bf r})\beta_{i}({\bf r})
\end{eqnarray}
with $X_{i}$ ($i=1,2,3$) the components of the vector $(X,Y,Z)$. 
The results for the $\tau^{2}\beta$ component (for the central region 
of the simulation) are summarized in Figure 10. The level of polarization 
in absolute temperature units is $\sim 1-2$ orders of magnitudes smaller 
than the $\propto\tau^{2}\Theta$ component. Convolving with the Planck 
HFI and a Gaussian beam profile with FWHM of 1', the peak values are 
$3$ nK and $25$nK, respectively, comparable to the level of the cosmological 
quadrupole polarization (section 3.1). As previously noted, both these 
components are frequency-independent, a fact that has obvious implications 
on the feasibility of separating these components out when the measurements 
will eventually reach the requisite sensitivity.

\begin{figure*}[t]
\centering
\vspace{9pt}
\epsfxsize=2.5in
\includegraphics[width=6.5cm,height=6cm]{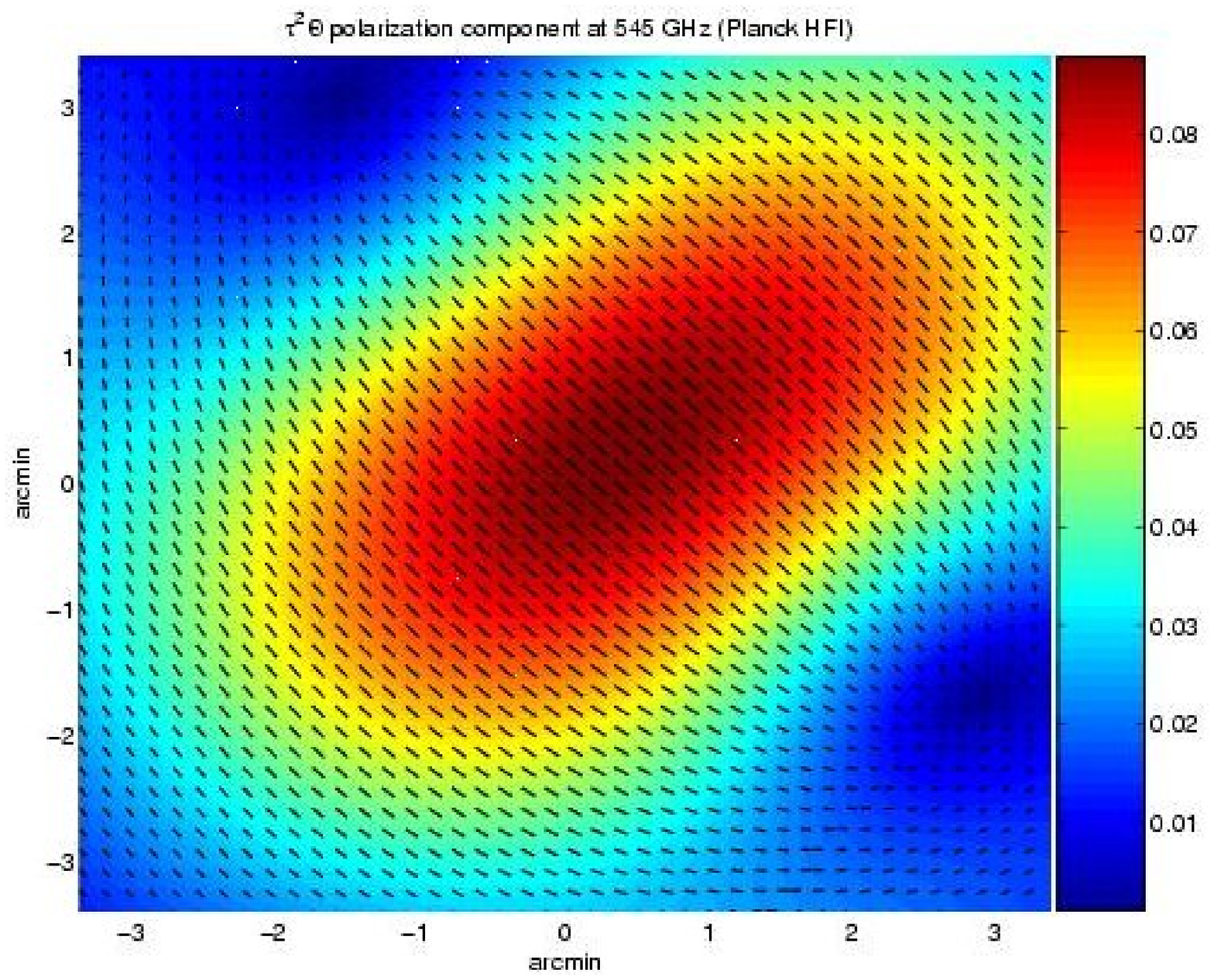}
\hspace{0.1in}
\includegraphics[width=6.5cm,height=6cm]{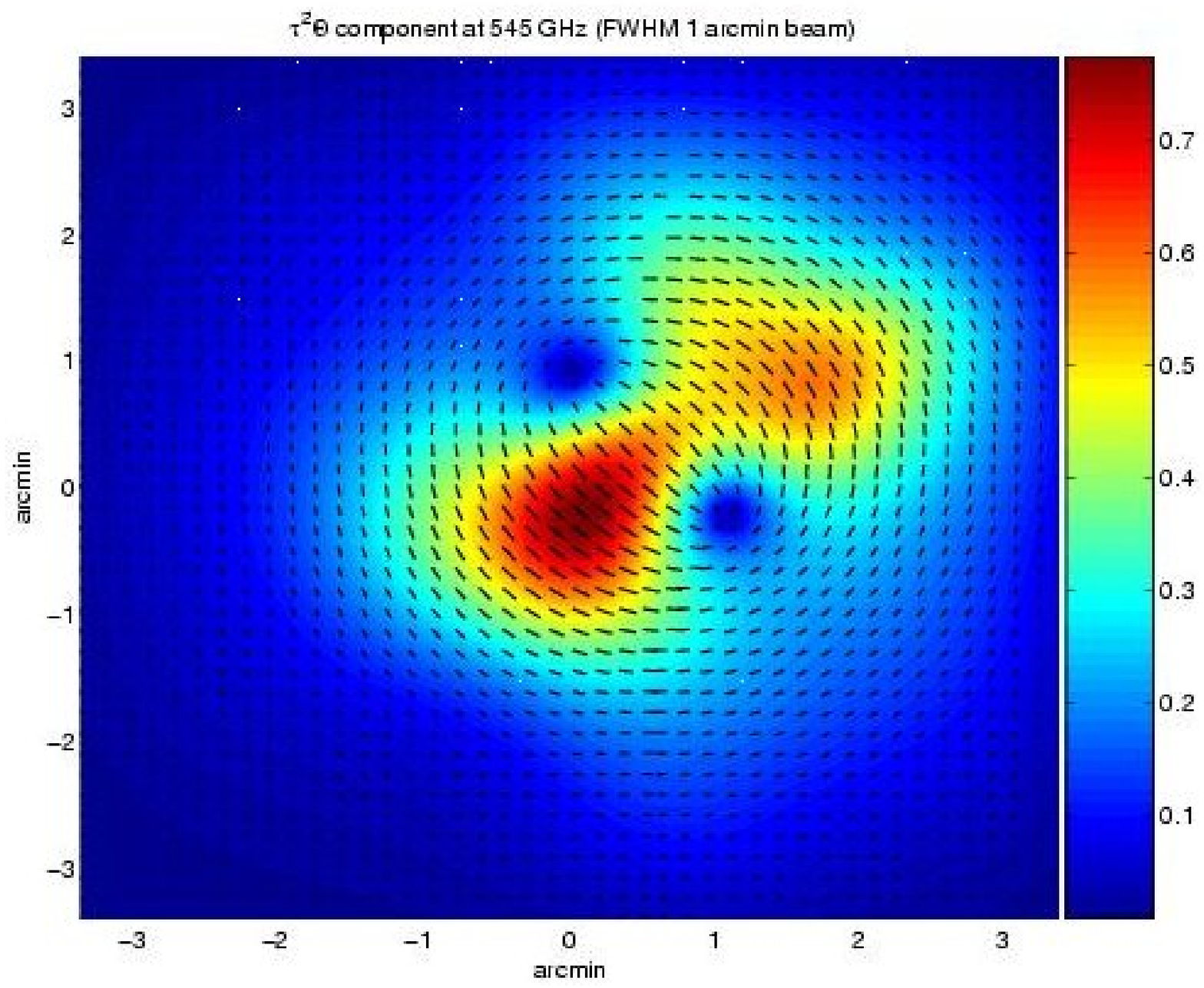}
\caption{The $\tau^{2}\Theta$ polarization component convolved with the 
Planck HFI 4.5' FWHM beam (left) and with a FWHM 1' beam (right). Color 
scale is in $\mu$K.}
\end{figure*}

\begin{figure*}[h]
\centering
\vspace{9pt}
\epsfxsize=2.5in
\includegraphics[width=6.5cm,height=6cm]{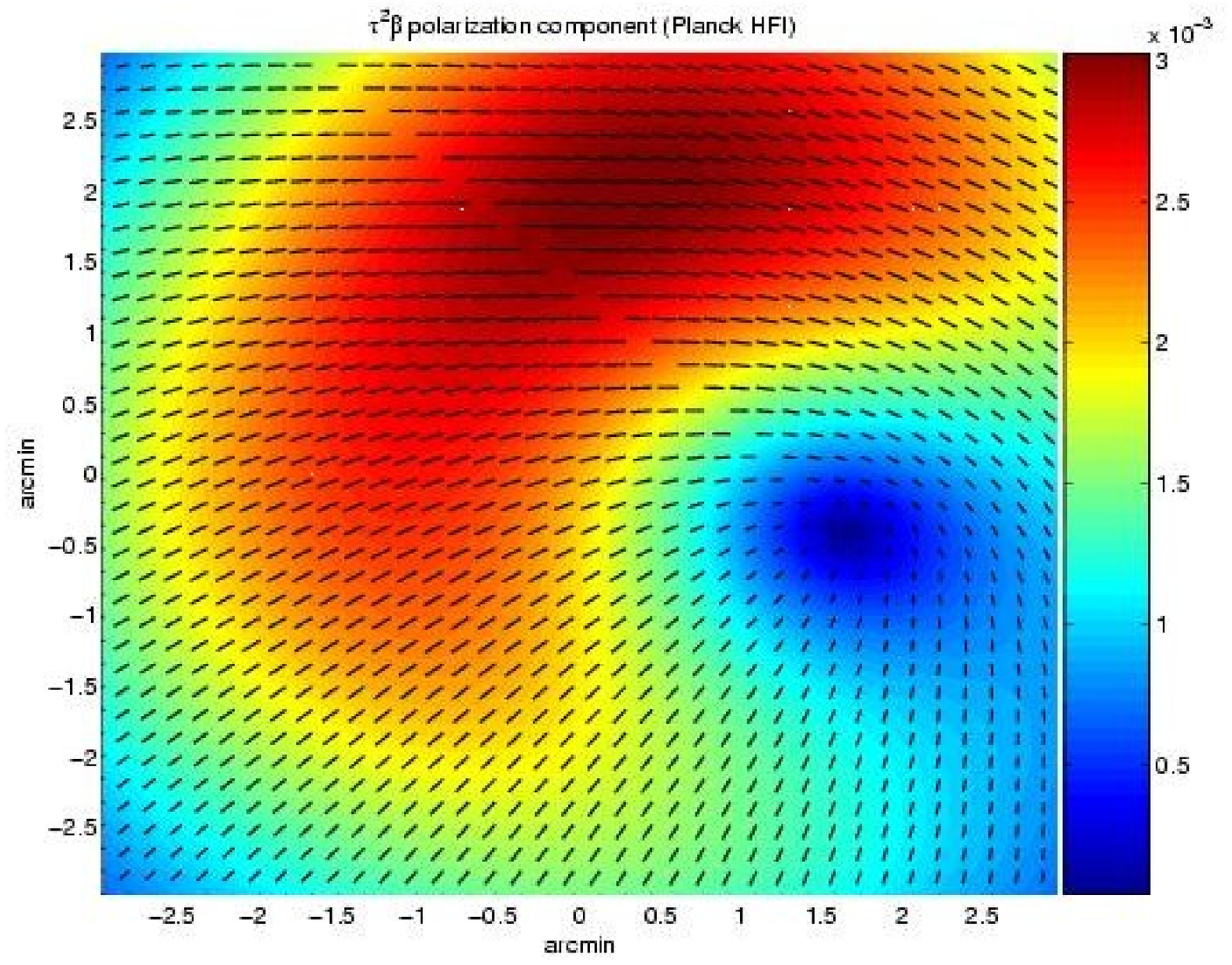}
\hspace{0.1in}
\includegraphics[width=6.5cm,height=6cm]{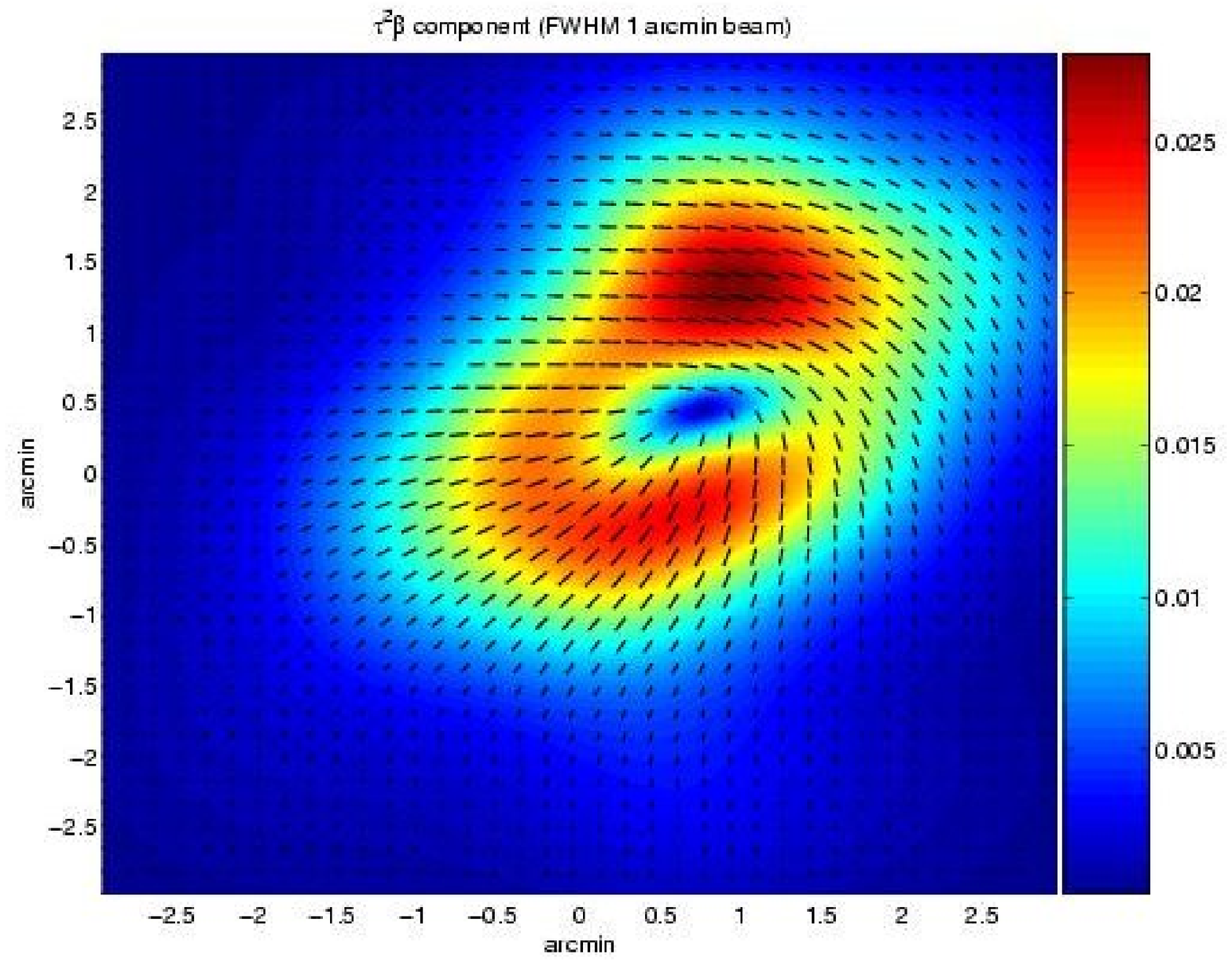}
\caption{The $\tau^{2}\beta$ polarization component convolved with the 
Planck HFI 4.5' FWHM beam (left) and with a FWHM 1' beam (right). Color 
scale is in $\mu$K.}
\end{figure*}

We end with a summary of the main result of this section: Most polarization 
signals induced by scattering of the CMB in clusters are substantially 
smaller than $1 \,\mu$K. An exception is the polarization induced by the 
double scattering thermal effect. Convolving the predictions for this 
component in the simulated cluster with a FWHM 1' beam, we find that the 
effective polarization signal is about $\sim 1 \,\mu K$. The much 
improved CMB observational capabilities - with projected sensitivities 
around $1$ $\mu$K and arcminute angular resolution - will make it feasible 
to measure the dominant polarization signals in clusters by upcoming 
experiments, thereby supplementing X-ray and (total intensity) S-Z 
measurements to probe IC gas properties and cluster morphology.
\bs

Acknowledgment: Work at Tel Aviv University was supported by Israel Science 
Foundation grant 225/03.

\end{document}